\documentclass[aps,prc,twocolumn,superscriptaddress,floatfix,amsmath,amssymb,nofootinbib,longbibliography]{revtex4-1}
\usepackage[utf8]{inputenc}
\usepackage[T1]{fontenc}

\usepackage{xcolor}
\usepackage{graphicx}

\usepackage{scalefnt}
\usepackage{times,txfonts}

\usepackage{amsmath}
\usepackage{amssymb}
\usepackage{amsfonts}
\usepackage{mathtools}

\usepackage{braket}

\usepackage[citecolor=blue,linkcolor=blue,urlcolor=blue]{hyperref}
\usepackage[ocgcolorlinks]{ocgx2}

\newcommand{\MeV}{\text{MeV}}
\newcommand{\fminv}{\text{fm}^{-1}}

\newcommand{\ai}{{\emph{ab initio}}}

\newcommand{\op}[2]{\ensuremath{{#1}^{(#2)}}}
\newcommand{\opno}[2]{\ensuremath{\tilde{#1}^{(#2)}}}
\newcommand{\mel}[2]{\ensuremath{{#1}_{#2}}}
\newcommand{\melno}[2]{\ensuremath{\tilde{#1}_{#2}}}
\newcommand{\crea}[1]{\ensuremath{a^{\dagger}_{#1}}}
\newcommand{\annih}[1]{\ensuremath{a_{#1}}}

\newcommand{\comm}[2]{\ensuremath{\left[#1,#2 \right]}}
\newcommand{\fcomm}[3]{\ensuremath{\left[#1,#2 \right]\rightarrow #3}}
\newcommand{\fcommtext}[3]{[\nobreak#1\nobreak,\nobreak\,\nobreak#2\nobreak]\nobreak$\nobreak\,\nobreak\rightarrow\nobreak\,$\nobreak#3}

\newcommand{\clebsch}[6]{C_{#1 #2 #3 #4}^{#5 #6}}

\newcommand{\sixj}[6]{\begingroup\setlength{\arraycolsep}{0.2em}\begin{Bmatrix} #1 & #2 & #3 \\ #4 & #5 & #6 \end{Bmatrix}\endgroup}

\allowdisplaybreaks

\begin{document}

\title{In-medium similarity renormalization group with three-body operators}
\author{M.~Heinz}
\email{mheinz@theorie.ikp.physik.tu-darmstadt.de}
\affiliation{Technische Universit\"at Darmstadt, Department of Physics, 64289 Darmstadt, Germany}
\affiliation{ExtreMe Matter Institute EMMI, GSI Helmholtzzentrum f\"ur Schwerionenforschung GmbH, 64291 Darmstadt, Germany}

\author{A.~Tichai}
\email{alexander.tichai@physik.tu-darmstadt.de}
\affiliation{Technische Universit\"at Darmstadt, Department of Physics, 64289 Darmstadt, Germany}
\affiliation{ExtreMe Matter Institute EMMI, GSI Helmholtzzentrum f\"ur Schwerionenforschung GmbH, 64291 Darmstadt, Germany}
\affiliation{Max-Planck-Institut f\"ur Kernphysik, Saupfercheckweg 1, 69117 Heidelberg, Germany}

\author{J.~Hoppe}
\email{jhoppe@theorie.ikp.physik.tu-darmstadt.de}
\affiliation{Technische Universit\"at Darmstadt, Department of Physics, 64289 Darmstadt, Germany}
\affiliation{ExtreMe Matter Institute EMMI, GSI Helmholtzzentrum f\"ur Schwerionenforschung GmbH, 64291 Darmstadt, Germany}

\author{K.~Hebeler}
\email{kai.hebeler@physik.tu-darmstadt.de}
\affiliation{Technische Universit\"at Darmstadt, Department of Physics, 64289 Darmstadt, Germany}
\affiliation{ExtreMe Matter Institute EMMI, GSI Helmholtzzentrum f\"ur Schwerionenforschung GmbH, 64291 Darmstadt, Germany}

\author{A.~Schwenk}
\email{schwenk@physik.tu-darmstadt.de}
\affiliation{Technische Universit\"at Darmstadt, Department of Physics, 64289 Darmstadt, Germany}
\affiliation{ExtreMe Matter Institute EMMI, GSI Helmholtzzentrum f\"ur Schwerionenforschung GmbH, 64291 Darmstadt, Germany}
\affiliation{Max-Planck-Institut f\"ur Kernphysik, Saupfercheckweg 1, 69117 Heidelberg, Germany}

\begin{abstract}
Over the past decade the in-medium similarity renormalization group (IMSRG) approach has proven to be a powerful and versatile \ai{} many-body method for studying medium-mass nuclei. So far, the IMSRG was limited to the approximation in which only up to two-body operators are incorporated in the renormalization group flow, referred to as the IMSRG(2). In this work, we extend the IMSRG(2) approach to fully include three-body operators yielding the IMSRG(3) approximation. We use a perturbative scaling analysis to estimate the importance of individual terms in this approximation and introduce  truncations that aim to approximate the IMSRG(3) at a lower computational cost. The IMSRG(3) is systematically benchmarked for different nuclear Hamiltonians for ${}^{4}\text{He}$ and ${}^{16}\text{O}$ in small model spaces. The IMSRG(3) systematically improves over the IMSRG(2) relative to exact results. Approximate IMSRG(3) truncations constructed based on computational cost are able to reproduce much of the systematic improvement offered by the full IMSRG(3). We also find that the approximate IMSRG(3) truncations behave consistently with expectations from our perturbative analysis, indicating that this strategy may also be used to systematically approximate the IMSRG(3).
\end{abstract}

\maketitle

\section{Introduction}
\label{sec:intro}

A key challenge in nuclear structure theory
is the calculation of the properties of atomic nuclei
with predictive power extending to unmeasured, exotic systems
targeted by modern rare-isotope facilities.
\emph{Ab initio} many-body approaches seek to accomplish this
by solving the many-body Schr{\"o}dinger equation in a systematically improvable manner
using two- and three-body nuclear interactions as input.
The rapid growth of the range of systems within reach of \ai{} many-body methods over the past two decades~\cite{Hebe15ARNPS,Morr17Tin,Hergert2020review}
can be understood in terms of improvements in the input interactions~\cite{Epel09RMP,Mach11PR,Hebe11fits,Ekst15sat,Ente17EMn4lo,Epel19nuclfFront,Hebe203NF,Jian20N2LOGO} and improvements in many-body approaches for medium-mass nuclei.

To access medium-mass and heavier systems,
the many-body approaches used in \ai{} calculations
start from an $A$-body reference state
on which corrections are systematically constructed.
These methods scale polynomially in the size of the computational basis $N$
rather than exponentially in the number of particles $A$, as is the case for the exact solution of the $A$-body Schr{\"o}dinger equation.
Examples of such methods are coupled-cluster (CC) theory~\cite{Hage14RPP,Bind14CCheavy},
the in-medium similarity renormalization group (IMSRG)~\cite{Tsuk11IMSRG,Herg16PR,Stroberg2019},
self-consistent Green's function (SCGF) theory~\cite{Dick04PPNP,Soma20SCGF}, and
many-body perturbation theory (MBPT)~\cite{Holt14Ca,Tich16HFMBPT,Tichai18BMBPT,Tichai2020review}.
These methods all share a many-body truncation
that can be systematically relaxed
and in the limit of no many-body truncation
recover the exact results.

In this work,
we focus on the systematic improvement of the IMSRG,
which is currently truncated at the normal-ordered two-body level,
the IMSRG(2) approximation.
In coupled-cluster theory,
many different methods have been developed
to approximately and exactly handle three-body effects
in many-body calculations~\cite{Lee84CCSDT1,Noga87CCSDT,Scus88CCSDT,Piec05CRCC,Taub08LambdaCCSD(T),Bind13expl3NLCCSD(T),Hage14RPP}.
These effects have been shown to be important for the reproduction of a range of observables,
such as $2^{+}$ excited-state energies at closed shells~\cite{Hage16Ni78}, dipole polarizabilities~\cite{Miorelli2018,Kauf2068Ni}, and nuclear $\beta$-decay matrix elements~\cite{Novario2020b}.
In these cases, the IMSRG(2) performance is deficient relative
to methods that are able to treat three-body effects~\cite{Simo17SatFinNuc}.
For the IMSRG,
truncations that include induced three-body effects have been applied to shell-model diagonalizations using universal shell-model interactions~\cite{Herg18IMSRGConfProc}.
In quantum chemistry,
the driven similarity renormalization group,
a similar many-body method to the IMSRG,
has been extended to approximately include three-body effects
in ways designed to reproduce the success of coupled-cluster theory in electronic systems~\cite{Li20DSRGthreebody}.
For the IMSRG, however, studies of the role of three-body operators for nuclear systems have not yet been performed.

To systematically study three-body operators in the IMSRG,
we extend the many-body truncation to the normal-ordered three-body level,
defined as the IMSRG(3) approximation, and construct various different approximate IMSRG(3) truncation schemes with reduced computational cost. We apply these truncation schemes to closed-shell systems in small model spaces and analyze their properties in detail using perturbative tools. We study how they compare to exact results obtained from full diagonalizations, analyze the systematics of the many-body expansion in these systems, and investigate how full IMSRG(3) results can be approximated at a lower computational cost.

In Sec.~\ref{sec:imsrg},
we give an overview of the IMSRG formalism.
Section~\ref{sec:imsrg3} discusses the IMSRG(3) truncation,
provides the fundamental commutators for the truncation,
gives an overview of the perturbative analysis to understand their relative importance,
and introduces approximate IMSRG(3) truncation schemes.
In Sec.~\ref{sec:applications},
we apply the IMSRG(3) and our approximate truncation schemes
to the closed-shell nuclei $^{4}$He and $^{16}$O.
Finally, we summarize our results in Sec.~\ref{sec:summary}.

\section{Many-body formalism}
\label{sec:imsrg}

\subsection{Operator representation}

In this work, an $A$-body operator
\begin{equation}\label{eq:free_space_operator}
    O = \op{O}{0} + \cdots + \op{O}{A}
\end{equation}
is composed of zero- through $A$-body parts,
given in second-quantized form by
\begin{equation}\label{eq:free_space_operator_second_quantized}
    \op{O}{A} = \frac{1}{{(A!)}^2}
    \sum_{p_1, \ldots, p_{2A}} \mel{O}{p_1 \cdots p_{2A}}
    \crea{p_{1}}\cdots \crea{p_{A}} \annih{p_{2A}} \cdots \annih{p_{A + 1}}
\end{equation}
with the antisymmetrized matrix elements $O_{p_1 \cdots p_{2A}}$
and the fermion creation (annihilation) operators
$\crea{p}$ ($\annih{p}$),
which create (annihilate) a particle in the single-particle state $\ket{p}$.

Normal-ordering techniques can be used to exactly rearrange $O$
into normal-ordered zero- through $A$-body parts,
\begin{equation}\label{eq:normal_ordered_operator}
    O = \opno{O}{0} + \cdots + \opno{O}{A},
\end{equation}
where the normal ordering is performed
with respect to a reference state
that is a good starting approximation for
the targeted ground or excited state.
The normal-ordered $A$-body parts are given by
\begin{equation}\label{eq:normal_ordered_operator_second_quantized}
    \opno{O}{A} = \frac{1}{{(A!)}^2}
    \sum_{p_1, \ldots, p_{2A}} \melno{O}{p_1 \cdots p_{2A}}
    :\crea{p_{1}}\cdots \crea{p_{A}} \annih{p_{2A}} \cdots \annih{p_{A + 1}}:.
\end{equation}
In Eqs.~\eqref{eq:normal_ordered_operator}
and~\eqref{eq:normal_ordered_operator_second_quantized},
the tilde distinguishes the normal-ordered operator
and its normal-ordered matrix elements
from their free-space equivalents
in Eqs.~\eqref{eq:free_space_operator}
and~\eqref{eq:free_space_operator_second_quantized}.
The normal ordering of the string of creation and annihilation operators
is indicated by the surrounding colons, $:\cdots:$.
In the following, we work exclusively with normal-ordered operators
and matrix elements
and leave the tilde off to simplify notation.

We focus on the case where the reference state
to describe an $A$-body system
is a single $A$-particle Slater determinant:
\begin{equation}
    \ket{\Phi} = \prod_{i = 1}^{A} \crea{p_i} \ket{0},
\end{equation}
where $\ket{0}$ is the vacuum,
the state where no particles are present.
For a single-particle state $\ket{p}$,
if it is occupied in the reference state,
then it has occupation number $n_p = 1$ and is called a hole state.
Similarly, if it is unoccupied in the reference state,
then it has $n_p = 0$ and is called a particle state.
The $A$-body Hilbert space is spanned by the reference state
and its elementary excitations
\begin{equation}
    \ket{\Phi_{i_1 \cdots i_B}^{a_1\cdots a_B}} =
    \crea{a_B} \cdots \crea{a_1} \annih{i_B} \cdots \annih{i_1} \ket{\Phi},
\end{equation}
which can be constructed by exciting the fermions in the hole states $\ket{i_1}$ through $\ket{i_B}$
into the particle states $\ket{a_1}$ through $\ket{a_B}$.
This state is a $B$-particle $B$-hole ($BpBh$) excited state,
where $B \leq A$.

A conventional notation for the normal-ordered Hamiltonian is
\begin{align}
    \phantom{H}
    & \begin{aligned}
    \mathllap{H} & = E + f + \Gamma + W
    \end{aligned} \\
    & \begin{aligned}
    & = E + \sum_{pq} f_{pq} :\crea{p} \annih{q}:
    + \frac{1}{(2!)^2}\sum_{pqrs} \Gamma_{pqrs} :\crea{p} \crea{q} \annih{s} \annih{r}: \\
    & \quad
    + \frac{1}{(3!)^2} \sum_{pqrstu} W_{pqrstu} :\crea{p} \crea{q} \crea{r} \annih{u} \annih{t} \annih{s}:\,,
    \end{aligned}
\end{align}
where $E$ is the reference-state expectation value
of the Hamiltonian, $\braket{\Phi | H | \Phi}$,
and $f$, $\Gamma$, and $W$ are
the normal-ordered one\nobreakdash-, two\nobreakdash-, and three-body parts
of the Hamiltonian.
For example, for a Hartree-Fock (HF) reference state,
$E$ is the Hartree-Fock energy, and $f$ is the Fock operator,
which is diagonal in the eigenbasis of the HF one-body density matrix.
The physical ground state of the system is not a single Slater determinant
but some linear combination of $\ket{\Phi}$ and its elementary excitations,
leading to an energy lower than the reference-state expectation value.
In the IMSRG and other many-body methods,
the task is to calculate the remaining correlation energy
beyond the Hartree-Fock level
to obtain the exact ground-state energy.

\subsection{In-medium similarity renormalization group}

The similarity renormalization group (SRG)~\cite{Wegn94SRG,Glaz93SRG,Bogn07SRG,Bogn10PPNP}
seeks to construct a continuous unitary transformation of the Hamiltonian
in the flow parameter $s$,
\begin{equation}
    H(s) = U(s) H U^{\dagger}(s)\,,
\end{equation}
which can be obtained by solving the flow equation
\begin{equation}
    \frac{dH(s)}{ds} = \comm{\eta(s)}{H(s)},
\end{equation}
where the initial condition is $H(s=0) = H$
and the choice of the anti-Hermitian generator $\eta(s)$
fixes the unitary transformation generated
over the course of the SRG evolution.

When $H(s)$ and $\eta(s)$ are vacuum normal ordered,
the ``free-space'' SRG evolution
of potentials can be used to reduce couplings between low and high momenta
for two- and three-nucleon potentials.
These ``softened'' potentials exhibit improved many-body convergence.
At the same time,
the evolution induces many-body forces,
a fact one can quickly verify by considering the commutator in second-quantized form.
The treatment of many-body interactions in the free-space SRG approach
is limited by the exponential cost of representing the $A$-body Hamiltonian
in a Jacobi or single-particle basis,
restricting this approach to the consistent evolution of two- and three-body forces~\cite{Jurg09SRG3N,Hebe12msSRG,Hebe203NF}.

In the IMSRG~\cite{Tsuk11IMSRG},
$H(s)$ and $\eta(s)$ are normal ordered with respect to $\ket{\Phi}$,
and the expression for the commutator
is brought into normal order using Wick's theorem~\cite{Wick50theorem}.
The in-medium normal ordering
captures many of the effects of induced many-body interactions,
which are always present in SRG evolutions,
through lower-body interactions of the normal-ordered Hamiltonian.
This is the feature that allows the IMSRG
to succeed for the solution of the many-body Schr{\"o}dinger equation for large systems
where the SRG quickly becomes computationally intractable.

The generator $\eta(s)$ in SRG applications
is typically chosen to decouple certain parts of
the Hamiltonian over the course of the evolution.
In the single-reference IMSRG,
$\eta(s)$ is chosen to suppress couplings
between the reference state and its elementary excitations~\cite{Herg16PR},
such that
\begin{equation}
    \braket{\Phi | H(s \rightarrow \infty) | \Phi_{i\cdots}^{a\cdots}} = 0\,.
\end{equation}
When this decoupling is achieved,
the unitary transformation generated by the IMSRG
is such that the matrix element $\braket{\Phi | H(s \rightarrow \infty) | \Phi} = E(s \rightarrow \infty)$ is the correlated energy of the state targeted by the reference state.

\subsection{Truncation schemes}

The IMSRG formalism is exact if one is able to keep track of all induced normal-ordered many-body contributions,
as it is simply a unitary transformation on the many-body Hamiltonian
that decouples the matrix element $\braket{\Phi | H | \Phi}$
from the remaining matrix elements.
For practical calculations,
the IMSRG solution must be restricted
to include only the operators up to some fixed particle rank.
The current standard truncation for nuclear structure applications is the IMSRG(2),
where all operators are truncated at the normal-ordered two-body level:
\begin{align}
    H(s) &= E(s) + f(s) + \Gamma(s)\,, \\
    \eta(s) &= \op{\eta}{1}(s) + \op{\eta}{2}(s)\,.
\end{align}

At this truncation, there are two approximations present. First, for Hamiltonians with three-body interactions, the residual normal-ordered three-body part of the Hamiltonian $W(s=0)$ is discarded,
which is the so-called normal-ordered two-body (NO2B) approximation~\cite{Hage07CC3N,Roth12NCSMCC3N}.
Second,
the commutator $\big[\op{\eta}{2}(s),\Gamma(s)\big]$
has a normal-ordered three-body part,
which is discarded in the IMSRG(2).
Some attempts to approximately capture the effects
of neglected induced three-body contributions in the IMSRG(2)
have been explored~\cite{Morr16imsrgphd},
but a systematic understanding
has not been formed.

The IMSRG(2) approximation has several desirable features
as a many-body method.
It scales polynomially
[specifically like $\mathcal{O}(N^6)$]
in the size of the single-particle basis $N$.
It is complete up to third order in MBPT,
but it is also nonperturbative
in that it resums $pp$/$hh$-ladder
and $ph$-ring diagrams~\cite{Herg16PR}.
Additionally, it is size extensive,
meaning that its error scales linearly in the size of the system.
This puts the IMSRG(2) in the same category
as many-body methods like CCSD~\cite{Hage14RPP} and ADC(3)~\cite{Cipo13Ox},
which are also nonperturbative and third-order complete
but differ from the IMSRG(2) in what higher-order MBPT contributions the methods include.

\section{IMSRG(3)}
\label{sec:imsrg3}

Extending the IMSRG
to the normal-ordered three-body level
yields the IMSRG(3) approximation.
The Hamiltonian and the generator now each have a normal-ordered three-body part,
\begin{align}
    H(s) &= E(s) + f(s) + \Gamma(s) + W(s)\,, \\
    \eta(s) &= \op{\eta}{1}(s) + \op{\eta}{2}(s) + \op{\eta}{3}(s)\,,
\end{align}
and this makes it possible to include the initial residual three-body interactions
exactly in the IMSRG(3) calculation.

\subsection{Fundamental commutators}

The IMSRG truncations are typically derived and implemented in terms
of the fundamental commutators of two many-body operators.
These fundamental commutators are the basic operations
that need to be evaluated in any IMSRG calculation.
For the commutator of a normal-ordered $K$-body operator $\op{A}{K}$
and a normal-ordered $L$-body operator $\op{B}{L}$,
the resulting operator has different normal-ordered $M$-body parts
$\op{C}{M}$:
\begin{equation}
    \comm{\op{A}{K}}{\op{B}{L}} = \sum_{M=|K-L|}^{K+L-1} \op{C}{M}.
\end{equation}
The fact that $M \leq K + L -1$ for the commutator
(as opposed to $M \leq K + L$ for a simple product of normal-ordered operators)
ensures that the many-body expansion is ``connected,''
which means that the IMSRG at any truncation level is size extensive.
We isolate the different $M$-body parts
that arise from the commutator of a $K$-body operator and an $L$-body operator,
using the following schematic notation in terms of their many-body ranks:
\begin{equation}
    \fcomm{K}{L}{M} \,.
\end{equation}

In the following, we provide the nonantisymmetrized expressions
for the matrix elements
of the fundamental commutators required by the IMSRG(3).
For the two- and three-body parts,
the matrix elements must be antisymmetrized by applying
the appropriate two- and three-body antisymmetrizer.
The expressions were derived using the automated normal-ordering tool
\textsc{drudge}~\cite{Zhao21drudgegit},
and, in cases where our expressions did not match those provided in Ref.~\cite{Herg16PR},
the results were verified by hand (see also the Appendices).

In the following sections,
the section headings employ the schematic notation introduced above,
where \fcommtext{$K$}{$L$}{$\circ$} is short for
\begin{equation}
    \comm{\op{A}{K}}{\op{B}{L}} \rightarrow C\,.
\end{equation}

\subsubsection{\texorpdfstring{$\fcomm{1}{1}{\circ}$}{11X commutators}}

\begin{align}
    \phantom{C_{12}}
    & \begin{aligned}
    \mathllap{C_{12}} &=
    \sum_{p}\left(A_{1p} B_{p2} - B_{1p} A_{p2}\right),
    \end{aligned}\label{eq:comm_111}\\
    & \begin{aligned}
    \mathllap{\op{C}{0}} &=
    \sum_{pq}(n_p \bar{n}_q - \bar{n}_p n_q) A_{pq} B_{qp}\,,
    \end{aligned}\label{eq:comm_110}
\end{align}
with $\bar{n}_p \equiv 1 - n_p$ and the one-body matrix elements of the result $C_{12}$.

\subsubsection{\texorpdfstring{$\fcomm{1}{2}{\circ}$}{12X commutators}}

\begin{align}
    \phantom{C_{1234}}
    & \begin{aligned}
    \mathllap{C_{1234}} &=
    2 \sum_{p}\left(A_{1p} B_{p234} - A_{p3} B_{12p4}\right),
    \end{aligned}\label{eq:comm_122}\\
    & \begin{aligned}
    \mathllap{C_{12}} &=
    \sum_{pq}(n_p \bar{n}_q - \bar{n}_p n_q) A_{pq} B_{1q2p}\,.
    \end{aligned}\label{eq:comm_121}
\end{align}

\subsubsection{\texorpdfstring{$\fcomm{2}{2}{\circ}$}{22X commutators}}

\begin{align}
    \phantom{C_{123456}}
    & \begin{aligned}
    \mathllap{C_{123456}} &=
    9 \sum_{p}\left(A_{3p45} B_{126p} - B_{3p45} A_{126p}\right),
    \end{aligned}\label{eq:comm_223}\\
    & \begin{aligned}
    \mathllap{C_{1234}} &=
    \frac{1}{2}\sum_{pq}(\bar{n}_p \bar{n}_q - n_p n_q) \\
    &\qquad \quad \times
    \left(A_{12pq} B_{pq34} - B_{12pq} A_{pq34}\right) \\
    &\quad - 4 \sum_{pq}(n_p \bar{n}_q - \bar{n}_p n_q) 
    A_{p23q} B_{1qp4} \,,
    \end{aligned}\label{eq:comm_222} \\
    & \begin{aligned}
    \mathllap{C_{12}} &=
    \frac{1}{2}\sum_{pqr}(\bar{n}_p \bar{n}_q n_{r} + n_p n_q \bar{n}_r) \\
    &\qquad \quad \times
    \left(A_{1rpq} B_{pq2r} - B_{1rpq} A_{pq2r}\right),
    \end{aligned}\label{eq:comm_221} \\
    & \begin{aligned}
    \mathllap{\op{C}{0}} &=
    \frac{1}{4} \sum_{pqrs}(n_p n_q \bar{n}_r \bar{n}_s - \bar{n}_p \bar{n}_q n_r n_s)
    A_{pqrs} B_{rspq}\,.
    \end{aligned}\label{eq:comm_220}
\end{align}

\subsubsection{\texorpdfstring{$\fcomm{1}{3}{\circ}$}{13X commutators}}

\begin{align}
    \phantom{C_{123456}}
    & \begin{aligned}
    \mathllap{C_{123456}} &=
    3 \sum_{p}\left(A_{3p} B_{12p456} - A_{p6} B_{12345p}\right),
    \end{aligned}\label{eq:comm_133}\\
    & \begin{aligned}
    \mathllap{C_{1234}} &=
    \sum_{pq}(n_p \bar{n}_q - \bar{n}_p n_q) A_{pq} B_{12q34p}\,.
    \end{aligned}\label{eq:comm_132}
\end{align}

\subsubsection{\texorpdfstring{$\fcomm{2}{3}{\circ}$}{23X commutators}}

\begin{align}
    \phantom{C_{123456}}
    & \begin{aligned}
    \mathllap{C_{123456}} &=
    \frac{3}{2} \sum_{pq}(\bar{n}_p \bar{n}_q - n_p n_q) \\
    & \qquad \quad \times
    \left(A_{12pq} B_{pq3456} - A_{pq45} B_{123pq6}\right) \\
    & \quad
    + 9 \sum_{pq}(\bar{n}_p n_q - n_p \bar{n}_q)
    A_{3pq6} B_{12q45p}\,,
    \end{aligned}\label{eq:comm_233}\\
    & \begin{aligned}
    \mathllap{C_{1234}} &=
    \sum_{pqr}(\bar{n}_p \bar{n}_q n_r + n_p n_q \bar{n}_r) \\
    &\quad \times 
    \left(A_{r1pq} B_{pq234r} - A_{pqr3} B_{12rpq4} \right),
    \end{aligned}\label{eq:comm_232}\\
    & \begin{aligned}
    \mathllap{C_{12}} &=
    \frac{1}{4} \sum_{pqrs}(n_p n_q \bar{n}_r \bar{n}_s - \bar{n}_p \bar{n}_q n_r n_s)
    A_{pqrs} B_{rs1pq2}\,.
    \end{aligned}\label{eq:comm_231}
\end{align}

\subsubsection{\texorpdfstring{$\fcomm{3}{3}{\circ}$}{33X commutators}}

\begin{align}
    \phantom{C_{123456}}
    & \begin{aligned}
    \mathllap{C_{123456}} &=
    \frac{1}{6} \sum_{pqr}(n_p n_q n_r + \bar{n}_p \bar{n}_q \bar{n}_r) \\
    & \qquad \quad \times
    \left(A_{123pqr} B_{pqr456} - B_{123pqr} A_{pqr456}\right) \\
    & \quad
    + \frac{9}{2} \sum_{pqr}(\bar{n}_p \bar{n}_q n_r + n_p n_q \bar{n}_r) \\
    & \qquad \quad \times
    \left(A_{pq345r} B_{12rpq6} - B_{pq345r} A_{12rpq6}\right),
    \end{aligned}\label{eq:comm_333}\\
    & \begin{aligned}
    \mathllap{C_{1234}} &=
    \frac{1}{6} \sum_{pqrs}(\bar{n}_p \bar{n}_q \bar{n}_r n_s - n_p n_q n_r \bar{n}_s) \\
    &\qquad \quad \times 
    \left(A_{12spqr} B_{pqr34s} - B_{12spqr} A_{pqr34s} \right)\\
    &\quad
    + \sum_{pqrs}(n_p n_q \bar{n}_r \bar{n}_s - \bar{n}_p \bar{n}_q n_r n_s)
    A_{pq1rs3} B_{rs2pq4}\,,
    \end{aligned}\label{eq:comm_332}\\
    & \begin{aligned}
    \mathllap{C_{12}} &=
    \frac{1}{12} \sum_{pqrst}(n_p n_q n_r \bar{n}_s \bar{n}_t + \bar{n}_p \bar{n}_q \bar{n}_r n_s n_t) \\
    & \quad \times
    \left(A_{st1pqr} B_{pqrst2} - B_{st1pqr} A_{pqrst2} \right),
    \end{aligned}\label{eq:comm_331} \\
    & \begin{aligned}
    \mathllap{\op{C}{0}} &=
    \frac{1}{36} \sum_{pqrstu}(n_p n_q n_r \bar{n}_s \bar{n}_t \bar{n}_u - \bar{n}_p \bar{n}_q \bar{n}_r n_s n_t n_u) \\
    & \quad \times
    A_{pqrstu} B_{stupqr}\,.
    \end{aligned}\label{eq:comm_330}
\end{align}

The computational cost of each commutator scales naively like
$\mathcal{O}(N^{K + L + M})$
in the size of the single-particle basis $N$.
As a result,
the cost of the full IMSRG(3) solution scales like
the cost of the \fcommtext{3}{3}{3} commutator,
$\mathcal{O}(N^9)$.
The full IMSRG(3) flow equations for the matrix elements
of the Hamiltonian are provided in Appendix~\ref{app:imsrg3_flow_equations}
along with a list of the differences between the expressions we provide and those given in Ref.~\cite{Herg16PR}.

\subsection{Generators}

In the IMSRG(3),
the extended many-body truncation introduces new matrix elements
of the Hamiltonian that couple the reference state and its excitations,
specifically $W_{ijkabc}$ and $W_{abcijk}$,
where $i$, $j$, and $k$ are hole-state indices
and $a$, $b$, and $c$ are particle-state indices.
Below we extend the standard generator definitions
used in the single-reference IMSRG(2)~\cite{Herg16PR}
to the three-body case,
seeking to suppress these matrix elements over the course of the evolution.

For the imaginary-time generator,
we choose the matrix elements
of the three-body part of the generator to be
\begin{subequations}
    \begin{align}
        \eta_{ijkabc} &= \text{sgn}(\Delta_{ijkabc}) W_{ijkabc}\,, \\
        \eta_{abcijk} &= \text{sgn}(\Delta_{abcijk}) W_{abcijk}\,,
    \end{align}
\end{subequations}
where we use the M{\o}ller-Plesset energy denominators
\begin{equation}\label{eq:mp_energy_deominators}
    \Delta_{ijkabc} = f_{ii} + f_{jj} + f_{kk} - (f_{aa} + f_{bb} + f_{cc}) = - \Delta_{abcijk}\,.
\end{equation}
Similarly, the matrix elements of the three-body White generator
are chosen to be
\begin{subequations}
    \begin{align}
        \eta_{ijkabc} &= \frac{W_{ijkabc}}{\Delta_{ijkabc}}\,, \\
        \eta_{abcijk} &= \frac{W_{abcijk}}{\Delta_{abcijk}}\,,
    \end{align}
\end{subequations}
and the matrix elements of the three-body arctan generator
are chosen to be
\begin{subequations}
    \begin{align}
        \eta_{ijkabc} &= \frac{1}{2}
        \arctan\left(\frac{2 W_{ijkabc}}{\Delta_{ijkabc}}\right), \\
        \eta_{abcijk} &= \frac{1}{2}
        \arctan\left(\frac{2 W_{abcijk}}{\Delta_{abcijk}}\right).
    \end{align}
\end{subequations}

\subsection{Perturbative analysis}\label{sec:perturbative_analysis}

In Ref.~\cite{Herg16PR},
a perturbative analysis of the IMSRG is presented
for the case where the NO2B approximation
and an HF reference state are used.
This analysis reveals the MBPT diagrammatic content
of the many-body method,
and we use it as a tool to understand
the contributions of different commutators in the IMSRG(3).
In the following we present the key ideas of the perturbative analysis
and refer the reader interested in a more formal treatment
to Ref.~\cite{Herg16PR}.

The connection from the IMSRG to MBPT is cleanly made when using the White generator,
with the matrix elements
\begin{subequations}
    \begin{align}
        \eta_{ia} &= \frac{f_{ia}}{\Delta_{ia}}\,, \\
        \eta_{ijab} &= \frac{\Gamma_{ijab}}{\Delta_{ijab}}\,, \\
        \eta_{ijkabc} &= \frac{W_{ijkaba}}{\Delta_{ijkabc}}\,,
    \end{align}
\end{subequations}
where $\Delta_{ia}$ and $\Delta_{ijab}$ are defined analogously to Eq.~\eqref{eq:mp_energy_deominators}.
Here and in the following $i$, $j$, and $k$ are hole single-particle indices,
and $a$, $b$, and $c$ are particle single-particle indices.
Using this generator,
the zero-body part of the IMSRG flow equations (up to the three-body level)
has three contributions from the \fcommtext{1}{1}{0}, \fcommtext{2}{2}{0}, and \fcommtext{3}{3}{0} commutators,
\begin{align}
    \phantom{\left(\frac{dE}{ds}\right)_{110}}
    & \begin{aligned}
    \mathllap{\left(\frac{dE}{ds}\right)_{110}} &= 
    \sum_{ia} \left(\eta_{ia}(s) f_{ai}(s) - \eta_{ai}(s) f_{ia}(s)\right) \\
    & = 2 \sum_{ia} \eta_{ia}(s) f_{ai}(s)
    = 2 \sum_{ia} \frac{f_{ia}(s) f_{ai}(s)}{\Delta_{ia}(s)}\,,
    \end{aligned}\label{eq:flow_eq110_part_pert} \\
    & \begin{aligned}
    \mathllap{\left(\frac{dE}{ds}\right)_{220}} &= 
    \frac{1}{2} \sum_{ijab} \eta_{ijab}(s) \Gamma_{abij}(s) \\
    & = \frac{1}{2} \sum_{ijab} \frac{\Gamma_{ijab}(s) \Gamma_{abij}(s)}{\Delta_{ijab}(s)}\,,
    \end{aligned}\label{eq:flow_eq220_part_pert} \\
    & \begin{aligned}
    \mathllap{\left(\frac{dE}{ds}\right)_{330}} &= 
    \frac{1}{18} \sum_{ijkabc} \eta_{ijkabc}(s) W_{abcijk}(s) \\
    & = \frac{1}{18} \sum_{ijkabc} \frac{W_{ijkabc}(s) W_{abcijk}(s)}{\Delta_{ijkabc}(s)}\,,
    \end{aligned}\label{eq:flow_eq330_part_pert}
\end{align}
which look remarkably similar to the second-order MBPT corrections to the energy.
Indeed, if one approximates the hole-particle block matrix elements
$f_{ia}(s)$, $\Gamma_{ijab}(s)$, and $W_{ijkabc}(s)$
by their basic suppression behavior due to the White generator~\cite{Herg16PR},
\begin{subequations}
    \begin{align}
        f_{ia}(s) &\approx f_{ia}(s=0) \exp(-s)\,, \\
        \Gamma_{ijab}(s) &\approx \Gamma_{ijab}(s=0) \exp(-s)\,, \\
        W_{ijkabc}(s) &\approx W_{ijkabc}(s=0) \exp(-s)\,,
    \end{align}
\end{subequations}
and one approximates the energy denominators by their initial values,
then Eqs.~\eqref{eq:flow_eq110_part_pert}--\eqref{eq:flow_eq330_part_pert}
can be analytically integrated to get the results
\begin{align}
    E(s\rightarrow\infty)_{110} &\approx \sum_{ia} \frac{f_{ia}(s=0) f_{ai}(s=0)}{\Delta_{ia}(s=0)}\,,\\
    E(s\rightarrow\infty)_{220} &\approx \frac{1}{4}\sum_{ijab}
    \frac{\Gamma_{ijab}(s=0) \Gamma_{abij}(s=0)}{\Delta_{ijab}(s=0)}\,,\\
    E(s\rightarrow\infty)_{330} &\approx \frac{1}{36} \sum_{ijkabc}
    \frac{W_{ijkabc}(s=0) W_{abcijk}(s=0)}{\Delta_{ijkabc}(s=0)}\,.
\end{align}
These are exactly the second-order MBPT corrections to the energy,
and this shows that these corrections are absorbed into the IMSRG correlation energy,
making the IMSRG at any many-body truncation second-order complete in MBPT
(as long as the matrix elements are able to be captured initially in the many-body truncation).

Extending this analysis to higher orders in MBPT
requires considering how the hole-particle matrix elements of $f$, $\Gamma$, and $W$
change over the course of the IMSRG evolution
beyond the basic suppression of their initial values.
On a high level, this corresponds to the IMSRG evolution
``dressing'' the one\nobreakdash-, two\nobreakdash-, and three-body vertices
with effective interaction contributions that generate higher-order MBPT diagrams.

To make this analysis systematic,
we focus on the case where we use an HF reference state
and work in the NO2B approximation,
where the initial off-diagonal matrix elements of $f$
and all the initial matrix elements of $W$ are 0.
Working with a M{\o}ller-Plesset MBPT partitioning of the initial Hamiltonian,
\begin{equation}
    H = f + g\, \Gamma\,,
\end{equation}
we have the following power-counting scheme:
\begin{align}
    f_{pp} &= \mathcal{O}(g^0)\,,\\
    \Gamma_{pqrs} &= \mathcal{O}(g^1)\,,
\end{align}
that is, the diagonal one-body matrix elements
are $\mathcal{O}(g^0)$
and the two-body matrix elements are $\mathcal{O}(g^1)$.
In this case, the hole-particle block of $f$
is induced by the \fcommtext{2}{2}{1} commutator
(the \fcommtext{1}{2}{1} commutator initially does not induce hole-particle contributions
because $f_{ia}$ and thus $\op{\eta}{1}$ are 0),
and the matrix elements of $W$ are induced by the \fcommtext{2}{2}{3} commutator.
This means
\begin{align}
    f_{ia} = \mathcal{O}(g^2)\,,\\
    W_{pqrstu} = \mathcal{O}(g^2)\,,
\end{align}
and, as a result, their contributions to the energy are both $\mathcal{O}(g^4)$.
\footnote{
This is true both for the direct flow into the energy via, for example,
the \fcommtext{3}{3}{0} commutator ($g^2 \times g^2)$
and for the indirect case via an induced two-body part from, for example,
the \fcommtext{2}{3}{2} commutator followed by the flow into the energy through the \fcommtext{2}{2}{0} commutator ($g^1 \times g^2 \times g^1$).
}

\begin{table}[t]
    \begin{ruledtabular}
    \begin{tabular}{ccc}

     Commutator & Cost & Perturbative order \\
    \hline
         $\fcomm{1}{1}{0}$ & $\mathcal{O}(N^2)$ & $g^4$ \\
         $\fcomm{1}{1}{1}$ & $\mathcal{O}(N^3)$ & $g^4$ \\
         $\fcomm{1}{2}{1}$ & $\mathcal{O}(N^4)$ & $g^5$ \\
         $\fcomm{1}{2}{2}$ & $\mathcal{O}(N^5)$ & $g^2$ \\
         $\fcomm{2}{2}{0}$ & $\mathcal{O}(N^4)$ & $g^2$ \\
         $\fcomm{2}{2}{1}$ & $\mathcal{O}(N^5)$ & $g^4$ \\
         $\fcomm{2}{2}{2}$ & $\mathcal{O}(N^6)$ & $g^3$ \\
         $\fcomm{2}{2}{3}$ & $\mathcal{O}(N^7)$ & $g^4$ \\
         $\fcomm{1}{3}{2}$ & $\mathcal{O}(N^6)$ & $g^5$ \\
         $\fcomm{1}{3}{3}$ & $\mathcal{O}(N^7)$ & $g^4$ \\
         $\fcomm{2}{3}{1}$ & $\mathcal{O}(N^6)$ & $g^5$ \\
         $\fcomm{2}{3}{2}$ & $\mathcal{O}(N^7)$ & $g^4$ \\
         $\fcomm{2}{3}{3}$ & $\mathcal{O}(N^8)$ & $g^5$ \\
         $\fcomm{3}{3}{0}$ & $\mathcal{O}(N^6)$ & $g^4$ \\
         $\fcomm{3}{3}{1}$ & $\mathcal{O}(N^7)$ & $g^6$ \\
         $\fcomm{3}{3}{2}$ & $\mathcal{O}(N^8)$ & $g^5$ \\
         $\fcomm{3}{3}{3}$ & $\mathcal{O}(N^9)$ & $g^6$ \\
    \end{tabular}
    \end{ruledtabular}
    \caption{
    The lowest-order perturbative contribution to the energy provided
    by each of the fundamental commutators along with their computational cost.
    }
    \label{tab:fundamental_commutators_pert_order}
\end{table}

Thus, the contribution of any induced two-body parts to $E$
is suppressed by $\mathcal{O}(g^1)$,
and the contributions of induced one- and three-body parts to $E$
are suppressed by $\mathcal{O}(g^2)$.
This allows one to quickly perturbatively estimate
the importance of different fundamental commutators,
provided in Table~\ref{tab:fundamental_commutators_pert_order}.

It is worth noting that the \fcommtext{1}{1}{1}, \fcommtext{1}{2}{2}, and \fcommtext{1}{3}{3} commutators
have higher perturbative importance than
their \fcommtext{1}{2}{1}, \fcommtext{2}{2}{2}, and \fcommtext{2}{3}{3} counterparts,
a consequence of the fact that they are sensitive
to the diagonal part of $f$, which is $\mathcal{O}(g^0)$.
The former \fcommtext{1}{$B$}{$B$} commutators
are responsible for the suppression
of the $B$-body hole-particle blocks of the Hamiltonian
and play a central role in the behavior of the IMSRG evolution.
This is intuitively similar to the central role the kinetic energy plays
in the free-space SRG.

A key result of the analysis in Ref.~\cite{Herg16PR} is that the IMSRG(2)
is complete up to third order in MBPT
and contains many fourth-order diagrams as well.
At the NO2B level, the IMSRG(3) accounts for the induced three-body effects,
which are what is missing for the complete inclusion of fourth-order diagrams in the IMSRG(2),
making the IMSRG(3) fourth-order complete (at the NO2B level)~\cite{Herg16PR}.

\subsection{Approximation schemes}\label{sec:approximation_schemes}

Due to the high computational cost of full IMSRG(3) calculations,
finding a way to approximate the IMSRG(3) truncation
would pave the way to large model-space IMSRG calculations
that approximately include the effects of three-body operators.
In the following,
we present approximation schemes
by including in each scheme
selected IMSRG(3) fundamental commutators
on top of the IMSRG(2).

The first major truncation beyond IMSRG(2)
we use includes the minimum commutators necessary to make
the truncation fourth-order complete in MBPT.
These are the \fcommtext{2}{2}{3}, \fcommtext{2}{3}{2}, \fcommtext{1}{3}{3}, and \fcommtext{3}{3}{0} commutators.
We refer to this truncation as
the IMSRG(3)-MP4 approximation.
The IMSRG(3)-MP4 is most similar to iterated coupled-cluster methods like CCSDT-1~\cite{Lee84CCSDT1,Watts95EOMCCSDT1,Hage14RPP}, as
both methods are fourth-order complete.
However, CCSDT-1 scales like $\mathcal{O}(A^3 N^4)$
[naively $\mathcal{O}(N^7)$, but $A$ is up to an order of magnitude smaller than $N$ in converged calculations],
while the IMSRG(3)-MP4 scales like $\mathcal{O}(N^7)$.

We note that in our studies we found that
the \fcommtext{2}{2}{3}, \fcommtext{2}{3}{2}, and \fcommtext{1}{3}{3} commutators are required at the NO2B level
for an approximate IMSRG(3) truncation
to include some three-body effects
and also be numerically stable.
Without the \fcommtext{2}{2}{3} and \fcommtext{2}{3}{2} commutators,
the zero- through two-body parts and the three-body part are decoupled,
and the results remain identical to the IMSRG(2) results.
Without the \fcommtext{1}{3}{3} commutator,
the induced three-body part is not properly suppressed over the course
of the evolution,
and the correlation energy does not seem to converge.

\begin{table*}[t]
    \begin{ruledtabular}
    \begin{tabular}{ccccccc}

    Commutator & Cost & \multicolumn{5}{c}{Included in \ldots} \\
     & & IMSRG(3)-MP4 & IMSRG(3)-$N^7$ & IMSRG(3)-$N^8$ & IMSRG(3)-$g^5$ & IMSRG(3) \\
    \hline
         $\fcomm{2}{2}{3}$ & $\mathcal{O}(N^7)$ & \checkmark & \checkmark & \checkmark & \checkmark & \checkmark \\
         $\fcomm{2}{3}{2}$ & $\mathcal{O}(N^7)$ & \checkmark & \checkmark & \checkmark & \checkmark & \checkmark \\
         $\fcomm{1}{3}{3}$ & $\mathcal{O}(N^7)$ & \checkmark & \checkmark & \checkmark & \checkmark & \checkmark \\
         $\fcomm{3}{3}{0}$ & $\mathcal{O}(N^6)$ & \checkmark & \checkmark & \checkmark & \checkmark & \checkmark \\
         $\fcomm{2}{3}{1}$ & $\mathcal{O}(N^6)$ &  & \checkmark & \checkmark & \checkmark & \checkmark \\
         $\fcomm{1}{3}{2}$ & $\mathcal{O}(N^6)$ &  & \checkmark & \checkmark & \checkmark & \checkmark \\
         $\fcomm{3}{3}{1}$ & $\mathcal{O}(N^7)$ &  & \checkmark & \checkmark & & \checkmark \\
         $\fcomm{2}{3}{3}$ & $\mathcal{O}(N^8)$ &  & & \checkmark & \checkmark & \checkmark \\
         $\fcomm{3}{3}{2}$ & $\mathcal{O}(N^8)$ &  & & \checkmark & \checkmark & \checkmark \\
         $\fcomm{3}{3}{3}$ & $\mathcal{O}(N^9)$ &  & &  &  & \checkmark
    \end{tabular}
    \end{ruledtabular}
    \caption{
    The computational cost of the IMSRG(3) fundamental commutators
    and whether they are included in various approximate and full IMSRG(3) truncation schemes.
    }
    \label{tab:approximation_schemes}
\end{table*}

Beyond the IMSRG(3)-MP4 truncation,
we consider two approaches to including further commutators.
The first is inclusion based on computational cost,
including first the cheapest of the remaining commutators
before including the more expensive commutators~\cite{Stro20triumftalk}.
The rationale here is that by using this approach
one can include as much ``physics'' as possible
while increasing the computational cost incrementally,
hopefully leading to a fairly faithful reproduction
of the full IMSRG(3) results.
The second approach is based on the perturbative analysis
discussed in Sec.~\ref{sec:perturbative_analysis},
where remaining commutators are included
in the order of their perturbative importance.
This physically motivated approach attempts
to capture as best as possible the available physics
in a consistent manner
before including ``higher-order'' effects.
One would hope to see that
these higher-order effects generate only small changes in energies
and in practical calculations
some ``complete'' lower-order approximation could be used.

Following the first approach,
including the \fcommtext{2}{3}{1}, \fcommtext{1}{3}{2}, and \fcommtext{3}{3}{1} commutators
on top of the IMSRG(3)-MP4 approximation
yields a truncation that includes all IMSRG(3) commutators
that cost $\mathcal{O}(N^7)$ or less.
We refer to this truncation as the IMSRG(3)-$N^7$ truncation.
The inclusion of the \fcommtext{2}{3}{3} and \fcommtext{3}{3}{2} commutators
on top of this truncation
yields the IMSRG(3)-$N^8$ truncation,
which includes all commutators that cost $\mathcal{O}(N^8)$ or less.
This truncation differs from the full IMSRG(3)
only by the missing \fcommtext{3}{3}{3} commutator.

Following the second approach,
we note that the IMSRG(3)-MP4 truncation already follows this approach,
including all of the IMSRG(3) commutators that are $\mathcal{O}(g^4)$ or less,
with the exception of the \fcommtext{1}{2}{1} commutator,
which is $\mathcal{O}(g^5)$ and is included in the IMSRG(2) truncation.
The next truncation we present includes the remaining $\mathcal{O}(g^5)$ commutators,
the \fcommtext{2}{3}{1}, \fcommtext{1}{3}{2}, \fcommtext{2}{3}{3}, and \fcommtext{3}{3}{2} commutators,
on top of the IMSRG(3)-MP4 truncation.
We refer to this truncation as the IMSRG(3)-$g^5$ truncation.
This truncation includes two commutators that cost $\mathcal{O}(N^8)$,
making that the cost of the truncation.
The two remaining commutators are $\mathcal{O}(g^6)$,
so this is the only complete perturbatively guided truncation
between the IMSRG(3)-MP4 and full IMSRG(3) truncations.

The inclusion of specific commutators in each of the approximate IMSRG(3)
truncation schemes discussed above is presented in Table~\ref{tab:approximation_schemes}.

\section{Applications}
\label{sec:applications}

In this section,
we investigate the IMSRG(3) truncation
and the approximate truncations discussed in Sec.~\ref{sec:approximation_schemes}
when applied to the closed-shell ${}^{4}\text{He}$ and ${}^{16}\text{O}$ using different nuclear Hamiltonians.

\subsection{Hamiltonians and basis sets}

For most of our calculations,
we focus on two sets of chiral Hamiltonians,
one using the N$^3$LO nucleon-nucleon ($NN$) potential from Ref.~\cite{Ente03EMN3LO}
SRG-evolved to a resolution scale $\lambda=1.8\:\fminv$,
which we refer to as the ``EM 1.8'' Hamiltonian,
and one using the ``EM 1.8/2.0'' potential from Ref.~\cite{Hebe11fits} with both $NN$ and three-nucleon ($3N$) interactions.
For the treatment of the three-body part of the $NN$+$3N$ Hamiltonian
when using the EM 1.8/2.0 potential,
we use the NO2B approximation~\cite{Hage07CC3N,Roth12NCSMCC3N}.

In Sec.~\ref{sec:hard_nn_results},
we explore how the trends seen for the 
soft EM 1.8 and EM 1.8/2.0 Hamiltonians
are affected by the choice of harder Hamiltonians.
We use three sets of $NN$-only Hamiltonians.
One uses the N$^3$LO $NN$ potential from Ref.~\cite{Ente03EMN3LO} (with no SRG evolution applied),
which we refer to as the ``EM 500'' Hamiltonian based on its regulator cutoff at $\Lambda=500\:\MeV$.
The other two use the N$^3$LO $NN$ potential from Ref.~\cite{Ente17EMn4lo}
with $\Lambda=450\:\MeV$ (referred to as the ``EMN 450'' Hamiltonian)
and $\Lambda=500\:\MeV$ (referred to as ``EMN 500'').

In addition,
we use reference states
constructed from different single-particle basis sets.
Our single-particle basis is characterized
by the maximum principal quantum number $e_\text{max} = (2n+l)_\text{max}$,
with the radial quantum number $n$
and the orbital angular momentum $l$.
In the simplest case,
we solve the spherically restricted HF equations
to obtain a variationally optimized HF solution.
Where an HF reference state is used,
the solution of the HF equations
and the solution of the IMSRG both take place in an $e_{\text{max}} = 2$ model space.
The HF calculations were performed using the solver from Ref.~\cite{Stro17imsrggit}.

As an alternative, we use so-called natural orbitals (NAT),
which are defined as the eigenstates of the one-body density matrix.
Following the prescription detailed in Ref.~\cite{Stra73nucldens},
the one-body density matrix is expanded up to second order in perturbation theory, 
which incorporates dynamic particle-hole correlation effects
in the construction of the single-particle basis,
leading to improved convergence properties
and reduced sensitivity to the underlying basis frequency~\cite{Tich19NatNCSM,Hopp2020natural}.
We follow the strategy of Ref.~\cite{Hopp2020natural},
where the one-body density matrix is constructed in a large model space with $e_\text{max}^\text{NAT}$.
Following the construction of the basis
and the transformation of the Hamiltonian matrix elements,
the basis and operators are truncated to a model space
with a smaller $e_{\text{max}}$,
which is used for the IMSRG solution.

When using $NN$-only Hamiltonians,
the construction of the NAT basis takes place
in an $e_\text{max}^\text{NAT}=14$ model space.
The basis and Hamiltonian are truncated to an $e_{\text{max}} = 2$ model space
for the following IMSRG calculation.
When using the EM 1.8/2.0 $NN$+$3N$ Hamiltonian,
the construction of the NAT basis takes place
in an $e_\text{max}^\text{NAT}=14$ model space
with an additional $E_{3,\text{max}} = 16 \geq e_1 + e_2 + e_3$ truncation
placed on the three-body matrix elements.
Again, the basis and Hamiltonian are truncated to an $e_{\text{max}} = 2$ model space
for the following IMSRG calculation.

The IMSRG calculations presented here all use the imaginary-time generator
and solve the IMSRG by directly integrating the flow equations
(as opposed to using the Magnus-expansion approach~\cite{Morr15Magnus}).
For the single-reference IMSRG(2),
it was found that the choice of the generator
(between the imaginary-time, White, and arctan generators)
only has a very small effect on the result of the IMSRG solution~\cite{Herg16PR}.
We also experimented with generator choice in the IMSRG(3) case
and found that choosing a different generator changed the results obtained for each truncation scheme by less than 1~keV,
an effect much smaller than the effects we discuss in the following sections.
It seems that the insensitivity to generator choice in the IMSRG(2)
extends also to the IMSRG(3).

\subsection{Helium-4}\label{sec:appl_he4}

\begin{figure*}[p!]
    \centering
    \includegraphics{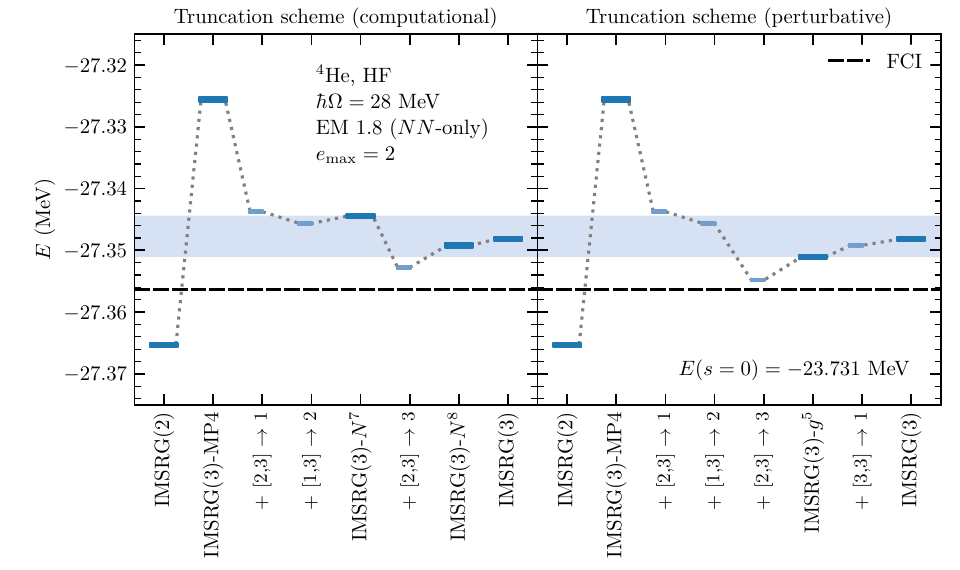}
    \caption{
        Ground-state energies of ${}^{4}\text{He}$
        obtained in various truncation schemes
        using the EM 1.8 $NN$-only Hamiltonian and an HF reference state
        following the computational (left panel)
        and perturbative (right panel) truncation ordering
        for the fundamental commutators.
        Thicker, darker bars correspond to the major truncations
        summarized in Table~\ref{tab:approximation_schemes}.
        Thinner, lighter bars
        correspond to intermediate truncations
        where a single fundamental commutator has been added
        relative to the truncation scheme to the left.
        The dashed line indicates the $e_{\text{max}}=2$ FCI result obtained for this Hamiltonian.
        The blue band indicates the range spanned
        by the results obtained from the IMSRG(3)-$N^7$ and IMSRG(3)-$g^5$ truncations.
        The starting HF energy is provided in the bottom right corner.
        \label{fig:he4_nn_only_hf_terms}
    }
\end{figure*}

In this section, we consider how the IMSRG solution for the ground-state energy of ${}^{4}\text{He}$
changes for different truncation schemes ranging from the IMSRG(2) to the full IMSRG(3) approximation.
We focus our discussion on the major truncations
discussed in Sec.~\ref{sec:approximation_schemes}
and presented succinctly in Table~\ref{tab:approximation_schemes}.
In the figures like Fig.~\ref{fig:he4_nn_only_hf_terms},
these truncations are visually indicated by the thicker bars.
We also introduce minor truncations,
which are defined as having one additional commutator included
relative to some previous truncation scheme.
For example,
one minor truncation scheme we consider
is the IMSRG(3)-$N^7$ $+$ \fcomm{2}{3}{3} truncation,
which has all $\mathcal{O}(N^7)$ commutators and the \fcommtext{2}{3}{3} commutator,
which is $\mathcal{O}(N^8)$.
The inclusion of the \fcommtext{3}{3}{2} commutator on top of this truncation
yields another major truncation, the IMSRG(3)-$N^8$ truncation.
These minor truncations are visually indicated by thinner bars.

We first focus on the case where we use the EM 1.8 $NN$-only Hamiltonian.
For the $NN$-only case,
we use an underlying oscillator frequency of $\hbar\Omega=28\:\MeV$,
which was determined by choosing the frequency
at which the ground-state energy that resulted from IMSRG(2) calculations using an HF reference state was minimal.
For comparison,
we provide exact results from the full configuration interaction (FCI) diagonalization of the $e_{\text{max}} = 2$ Hamiltonian.
In the absence of a many-body truncation,
this would be the exact result the IMSRG would be able to obtain,
and comparing against this result for different approximations
allows us to gain insight into the effect of the many-body truncations at play.

In Fig.~\ref{fig:he4_nn_only_hf_terms},
we show the ground-state energies for ${}^{4}\text{He}$
obtained using different IMSRG truncation schemes
using the EM 1.8 $NN$-only Hamiltonian
and an HF reference state.
In both panels,
we start from the IMSRG(2) truncation
and add commutators until we reach the IMSRG(3) truncation on the right.

In the left panel of Fig.~\ref{fig:he4_nn_only_hf_terms},
we follow the computational approach to organizing the IMSRG(3) fundamental commutators.
At the IMSRG(2)-truncation level,
the ground-state energy only differs from the FCI result by 9~keV.
The first truncation we consider beyond the IMSRG(2) is always the IMSRG(3)-MP4 truncation,
which in all systems we investigated delivered a sizable repulsive correction to the energy.
This is consistent with our understanding of the diagrammatic content of the IMSRG(2)
and the nature of the missing fourth-order MBPT energy corrections.
The inclusion of fundamental commutators up to the IMSRG(3)-$N^7$ truncation
brings the correlated energy back down towards the FCI result.
The next two commutators that are included in the IMSRG(3)-$N^8$ truncation
provide significant contributions that partially cancel.
The size of their individual contributions can be understood
by the fact that they are both fifth-order [$\mathcal{O}(g^5)$] in our perturbative counting
[to be compared with the $\mathcal{O}(g^6)$ contribution of \fcommtext{3}{3}{1}, which is the final commutator that contributes to the IMSRG(3)-$N^7$].
The contribution of the \fcommtext{3}{3}{3} commutator
to arrive at the full IMSRG(3) truncation is small,
and the final IMSRG(3) ground-state energy differs from the FCI result by 8~keV.

In the right panel,
we show the same information for the case where the perturbative ordering
of fundamental commutators is used.
We see that the $\mathcal{O}(g^5)$ commutators added from the IMSRG(3)-MP4 truncation to the IMSRG(3)-$g^5$ truncation
deliver contributions to the energy that are
generally smaller than the fourth-order shift between IMSRG(2) and IMSRG(3)-MP4 truncations
and generally larger than the sixth-order shifts between the IMSRG(3)-$g^5$ and the IMSRG(3) truncations,
which is consistent with the perturbative counting.

When discussing the contributions of commutators,
it is worth noting that the contribution of an added commutator to the energy
also depends on which other commutators are also included in that truncation.
In this context,
the one-by-one inclusion of fundamental commutators formally does not commute.
In practice, however,
we see that the size of the contribution of a specific commutator
is not strongly sensitive to the order in which it is included relative to other commutators.
One can see this behavior when comparing the two panels of Fig.~\ref{fig:he4_nn_only_hf_terms}.
Of course, substantial rearrangement of the commutators
(in particular, changing the order of two commutators that give large contributions to the energy)
can change this picture.
Our discussion, however, is built around the major truncation schemes
discussed in Sec.~\ref{sec:approximation_schemes},
restricting the freedom we have to move commutators around in between.
As far as we have seen in our explorations,
the quasiadditive nature of the inclusion of commutators and their energy contributions
seems to qualitatively hold within these restrictions.

\begin{figure*}[p!]
    \centering
    \includegraphics{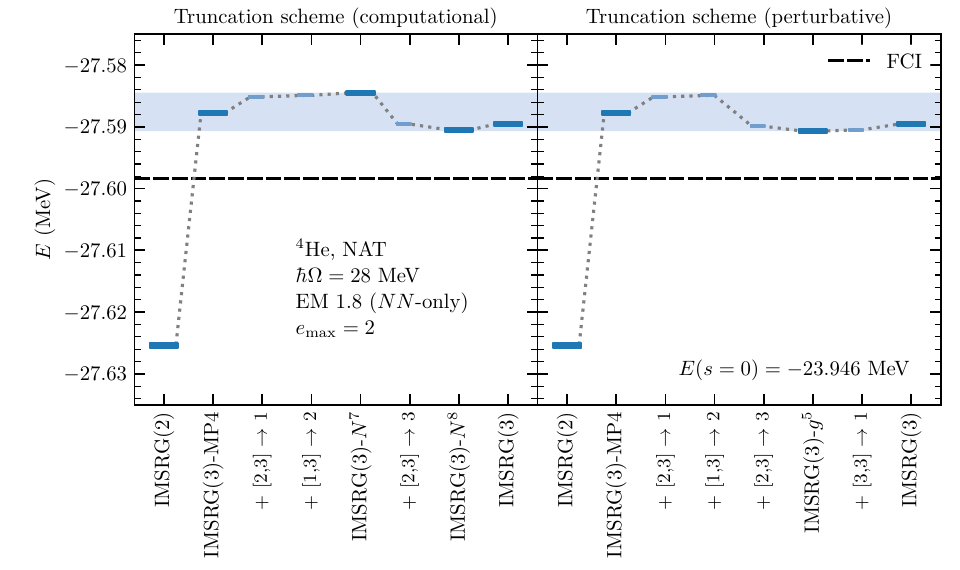}
    \caption{
        Same as Fig.~\ref{fig:he4_nn_only_hf_terms}
        but using a NAT reference state.
        \label{fig:he4_nn_only_nat_terms}
    }
\end{figure*}

In Fig.~\ref{fig:he4_nn_only_nat_terms},
we present results
for ${}^{4}\text{He}$ when using the EM 1.8 $NN$-only Hamiltonian
and a NAT reference state.
The same oscillator frequency is used as for the $NN$-only HF case
($\hbar \Omega = 28\:\MeV$).
The IMSRG(2) error to the FCI result is in this case 27~keV.
In the left panel,
following the repulsive IMSRG(3)-MP4 corrections to the energy,
we see that the commutators added to give the IMSRG(3)-$N^7$
give additional small repulsive shifts to the energy.
The $\mathcal{O}(N^8)$ commutators give slightly larger attractive contributions,
and the \fcommtext{3}{3}{3} commutator again delivers a very small contribution.
The final IMSRG(3) energy differs from the FCI result by 9~keV.
This is a considerable improvement over the IMSRG(2) result,
although all of the results discussed here
are quite good (sub-1\% error)
when compared to the total ground-state energy or the correlation energy.

In the right panel,
we see that the general size of energy contributions
follows the perturbative counting.
The size of all contributions beyond the IMSRG(3)-MP4 truncation
is substantially smaller than in the HF case discussed previously
(note that the relative scale on the energy in the graph is identical
in Figs.~\ref{fig:he4_nn_only_hf_terms} and~\ref{fig:he4_nn_only_nat_terms}).
In particular,
because the sixth-order commutator contributions are so small,
the IMSRG(3)-$g^5$ approximates the full IMSRG(3) extremely well.

\begin{figure}[t!]
    \centering
    \includegraphics{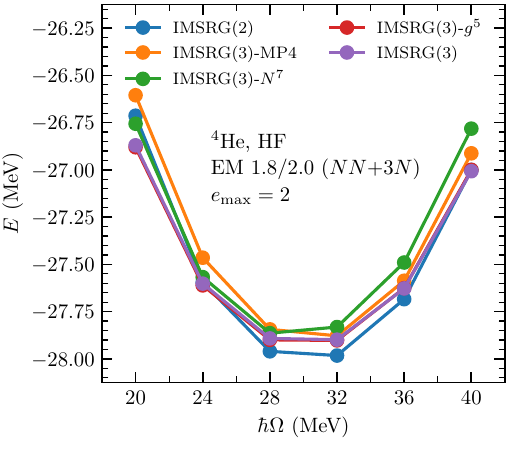}
    \caption{
        Ground-state energies of ${}^{4}\text{He}$
        using the EM 1.8/2.0 Hamiltonian
        and an HF reference state
        obtained in several IMSRG truncation schemes
        at a broad range of frequencies.
        \label{fig:he4_nn_and_3n_hf_frequency_scan}
    }
\end{figure}

Now we switch our focus to the case
where we use the EM 1.8/2.0 $NN$+$3N$ Hamiltonian.
We investigated the oscillator frequency sensitivity of the IMSRG(3) truncations
in ${}^{4}\text{He}$ using an HF reference state.
This system exhibits substantial frequency dependence
because $NN$+$3N$ Hamiltonians tend to give greater frequency dependence than their $NN$-only counterparts
and the HF basis depends more strongly on the frequency than the NAT basis.
This is because the NAT basis seeks to reduce frequency dependence by construction.

In Fig.~\ref{fig:he4_nn_and_3n_hf_frequency_scan},
we show the ground-state energy
obtained using several IMSRG truncations ranging from the IMSRG(2) to the IMSRG(3)
for a broad range of oscillator frequencies.
Generally,
we find that the results for the different truncations
remain quite close together (within a spread of 300~keV)
even as the energy varies over a range of 1.5~MeV.
This suggests that the variance in the energy
is entirely due to harsh infrared and ultraviolet cutoffs imposed by the $e_{\text{max}}=2$ model space
and not due to the many-body truncations,
which would be improved by the IMSRG(3).
It is of course possible that in calculations with larger model spaces
one might see systematic differences in the frequency dependence
of the energy resulting from different IMSRG truncations.

A couple systematic trends can be identified in Fig.~\ref{fig:he4_nn_and_3n_hf_frequency_scan}.
First, the IMSRG(3)-MP4 provides a repulsive contribution on top of the IMSRG(2) at all frequencies.
Second, the IMSRG(3)-$g^5$ and IMSRG(3) lines lie basically on top of each other,
indicating that the IMSRG(3)-$g^5$ reliably approximates the IMSRG(3).
The same cannot be said for the IMSRG(3)-$N^7$.
Finally, the IMSRG(3) results always lie below the IMSRG(3)-MP4 results.

\begin{figure}[t!]
    \centering
    \includegraphics{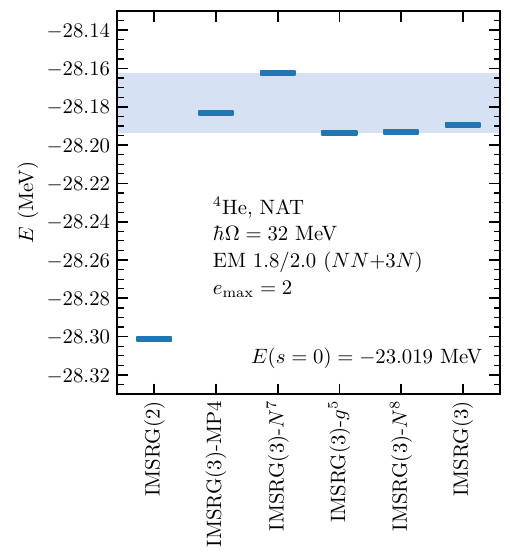}
    \caption{
        Ground-state energies of ${}^{4}\text{He}$
        obtained in various truncation schemes
        using the EM 1.8/2.0 Hamiltonian
        and a NAT reference state.
        The blue band indicates the range spanned
        by the results obtained from the IMSRG(3)-$N^7$ and IMSRG(3)-$g^5$ truncations.
        The starting energy of the NAT reference state is provided in the bottom right corner.
        \label{fig:he4_nn_and_3n_nat_major_truncs}
    }
\end{figure}

In Fig.~\ref{fig:he4_nn_and_3n_nat_major_truncs},
we present the ${}^{4}\text{He}$ ground-state energies
obtained in various IMSRG truncation schemes
using the EM 1.8/2.0 Hamiltonian
and a NAT reference state.
The oscillator frequency of $\hbar \Omega = 32\:\MeV$
was determined by choosing the frequency
at which the HF IMSRG(2) energy result was minimal for this Hamiltonian
(see Fig.~\ref{fig:he4_nn_and_3n_hf_frequency_scan}).
Overall, the corrections offered by approximate IMSRG(3) truncations are larger in magnitude
than in the $NN$-only case,
with the IMSRG(2) and IMSRG(3) results differing by 112~keV
(compare with the difference of 36~keV in the $NN$-only case).
We see similar trends as in the $NN$-only case,
with a large repulsive correction from the IMSRG(3)-MP4 truncation
and a smaller repulsive correction from the IMSRG(3)-$N^7$.
The $\mathcal{O}(N^8)$ fifth-order commutators provide attractive corrections,
and the final IMSRG(3) result lands between the IMSRG(3)-$N^7$ and IMSRG(3)-$g^5$ results,
as indicated by the blue band.

\subsection{Oxygen-16}\label{sec:appl_o16}

\begin{figure*}[p!]
    \centering
    \includegraphics{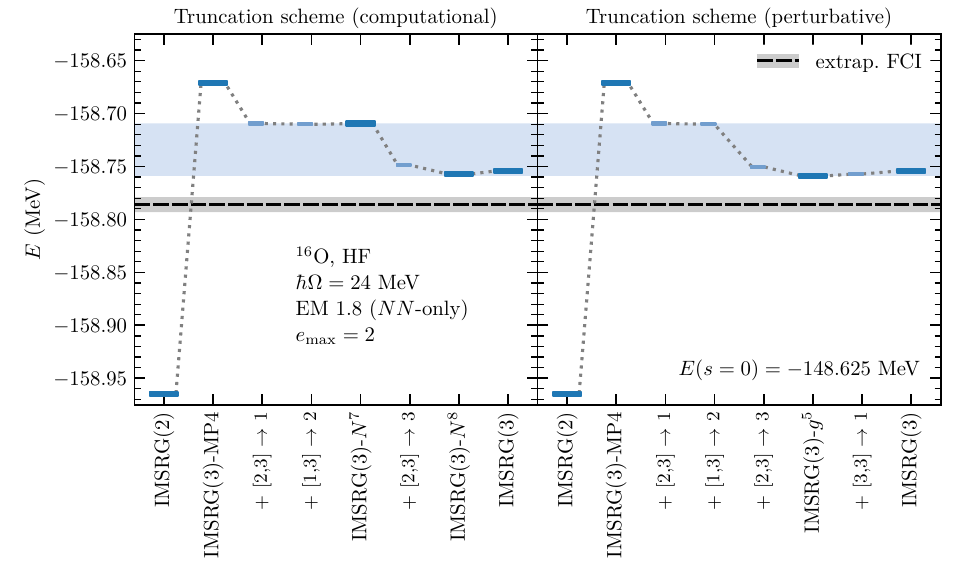}
    \caption{
        Ground-state energies of ${}^{16}\text{O}$
        obtained in various truncation schemes
        using the EM 1.8 $NN$-only Hamiltonian and an HF reference state
        following the computational (left panel)
        and perturbative (right panel) truncation ordering
        for the fundamental commutators.
        Thicker, darker bars correspond to the major truncations
        summarized in Table~\ref{tab:approximation_schemes}.
        Thinner, lighter bars
        correspond to intermediate truncations
        where a single fundamental commutator has been added
        relative to the truncation scheme to the left.
        The dashed line indicates the $e_{\text{max}}=2$ extrapolated FCI result obtained for this Hamiltonian (see main text for details).
        The blue band indicates the range spanned
        by the results obtained from the IMSRG(3)-$N^7$ and IMSRG(3)-$g^5$ truncations.
        The starting HF energy is provided in the bottom right corner.
        \label{fig:o16_nn_only_hf_terms}
    }
\end{figure*}

In this section,
we consider the IMSRG solution for the ground-state energy of ${}^{16}\text{O}$.
We first focus on the case where we use the EM 1.8 $NN$-only Hamiltonian.
In this case,
we use an oscillator frequency of $\hbar \Omega=24\:\MeV$.
For $NN$-only results,
we provide for comparison
extrapolated FCI results.
These results were obtained from a series of CI calculations
with increasing $N_{\text{max}}$
(the model space truncation for the approach)
from 0 to 8
using the \textsc{kshell} code~\cite{Shim19Kshell}.
The results from $N_{\text{max}}=2$ to 8 were then fit to an exponential function
to obtain the $N_{\text{max}}\rightarrow \infty$ extrapolated value~\cite{Roth11SRG}.
The uncertainty in the extrapolation was assessed by leaving
out one of the $N_{\text{max}}=2$, 4, 6 points
and fitting the exponential to the remaining three points
(the highest-quality $N_{\text{max}}=8$ point was always included).
The largest deviation from the full fit value and the subsampled fit values
is taken to be the uncertainty.

In Fig.~\ref{fig:o16_nn_only_hf_terms},
we show the ground-state energies of ${}^{16}\text{O}$
as obtained from different truncation schemes
when using an HF reference state.
The IMSRG(2) result differs from the exact result
by about 180~keV,
which corresponds to an error of 1.8\% in the correlation energy.
The IMSRG(3)-MP4 approximation provides a large, repulsive correction
to the IMSRG(2) result.
In the left panel,
we see that the \fcommtext{2}{3}{1} commutator included in the IMSRG(3)-$N^7$ truncation
provides a small, but significant attractive correction
and the \fcommtext{2}{3}{3} commutator included in the IMSRG(3)-$N^8$
delivers most of the remaining attraction
needed to produce the IMSRG(3) result.
The final IMSRG(3) result differs from the extrapolated FCI result
by only 32~keV,
which corresponds to an error of about 0.3\% in the correlation energy.
In the right panel,
we see that the perturbative counting of commutators
continues to be predictive,
with the smallest contributions belonging to the sixth-order commutators.
As a result, the IMSRG(3)-$g^5$ result lies quite close to the IMSRG(3) result.

\begin{figure*}[p!]
    \centering
    \includegraphics{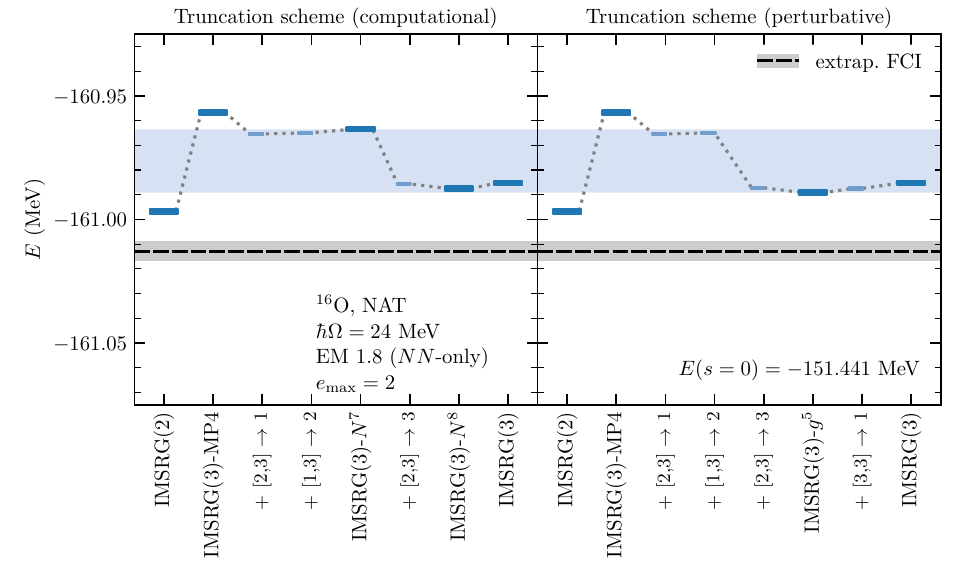}
    \caption{
        Same as Fig.~\ref{fig:o16_nn_only_hf_terms}
        but using a NAT reference state.
        \label{fig:o16_nn_only_nat_terms}
    }
\end{figure*}

In Fig.~\ref{fig:o16_nn_only_nat_terms},
we switch to a NAT reference state,
still considering ${}^{16}\text{O}$ using the EM 1.8 $NN$-only Hamiltonian.
The difference between the IMSRG(2) result
and the exact result is only 16~keV,
making the IMSRG(2) result in this case remarkably good.
The correction provided by the IMSRG(3)-MP4 truncation
is still repulsive, but considerably smaller than in the HF case.
In the left panel,
we see that again the \fcommtext{2}{3}{1} and \fcommtext{2}{3}{3} commutators
deliver the main contributions to corrections
provided by the IMSRG(3)-$N^7$ and IMSRG(3)-$N^8$ truncations, respectively.
The final IMSRG(3) result differs from the extrapolated FCI result
by 28~keV,
quite similar to the difference in the HF case.
The right panel shows that convergence to the IMSRG(3) result
in the perturbative counting approach is systematic in this case as well.

\begin{figure}[t!]
    \centering
    \includegraphics{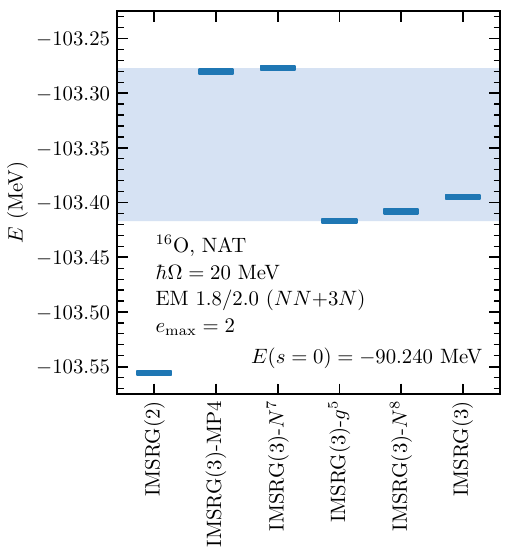}
    \caption{
        Ground-state energies of ${}^{16}\text{O}$
        obtained in various truncation schemes
        using the EM 1.8/2.0 Hamiltonian
        and a NAT reference state.
        The blue band indicates the range spanned
        by the results obtained from the IMSRG(3)-$N^7$ and IMSRG(3)-$g^5$ truncations.
        The starting energy of the NAT reference state is provided in the bottom right corner.
        \label{fig:o16_nn_and_3n_nat_major_truncs}
    }
\end{figure}

Switching to the EM 1.8/2.0 Hamiltonian,
we consider in Fig.~\ref{fig:o16_nn_and_3n_nat_major_truncs} the IMSRG solution for various truncations
for ${}^{16}\text{O}$ using a NAT reference state,
where the underlying oscillator frequency is $\hbar \Omega = 20\:\MeV$.
In this case,
the IMSRG(3)-MP4 truncation result is about 270~keV
more repulsive than the IMSRG(2) result,
and the IMSRG(3)-$N^7$ provides only small corrections to the IMSRG(3)-MP4 result.
These results differ substantially from those obtained from the remaining truncation schemes,
which contain all the
$\mathcal{O}(N^8)$ fifth-order commutators.
Of the systems we studied,
this is the system with the largest contribution by these commutators,
making the IMSRG(3)-$g^5$, for example, a substantial improvement over the IMSRG(3)-$N^7$
due to its inclusion of these higher-cost fifth-order commutators
that are neglected in the IMSRG(3)-$N^7$.
We see that again the large band resulting from the IMSRG(3)-$N^7$ and IMSRG(3)-$g^5$ results includes the IMSRG(3) result.

\subsection{Analysis of truncation performance}
\label{sec:truncation_performance}

\begin{figure}[t!]
    \centering
    \includegraphics{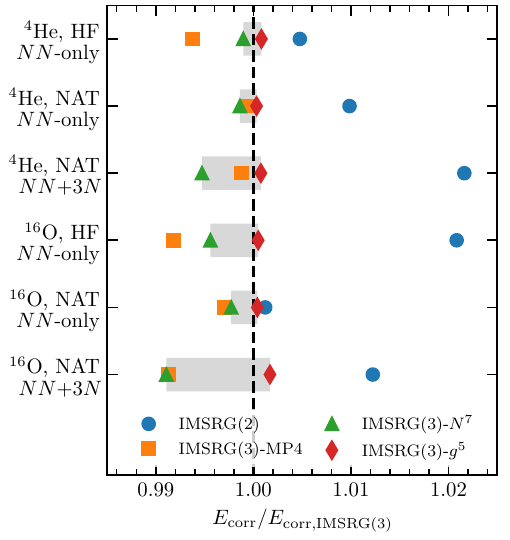}
    \caption{
        Ratios of correlation energies
        obtained in IMSRG(2) and approximate IMSRG(3) calculations
        relative to the IMSRG(3) correlation energies
        for different systems discussed in Secs.~\ref{sec:appl_he4} and~\ref{sec:appl_o16}.
        The gray band indicates the range spanned by
        the IMSRG(3)-$N^7$ and IMSRG(3)-$g^5$ results.
        \label{fig:truncation_summary}
    }
\end{figure}

Next,
we consider the relative performance of the different IMSRG truncations
over all systems considered.
These trends are summarized in Fig.~\ref{fig:truncation_summary}.
In this figure,
we compare the correlation energy,
defined as
\begin{equation}
    E_{\text{corr}} = E(s\rightarrow \infty) - E(s=0)\,,
\end{equation}
for the IMSRG(2) and approximate IMSRG(3) truncations
relative to the IMSRG(3) correlation energy.
The vertical line at $x=1.0$ indicates the IMSRG(3) correlation energy.
In the previous sections,
we saw that in most cases the IMSRG(3) energies
were closer to the exact results
obtained via FCI and extrapolated FCI calculations
(with the exception of the ${}^{16}\text{O}$ case
with the EM 1.8 $NN$-only Hamiltonian and the NAT reference state).
This intuitively matches the expected behavior of the many-body expansion,
where including higher many-body ranks in the many-body expansion
allows the truncated methods to systematically approach the exact result.
In this figure and the following discussion,
we frame things relative to the IMSRG(3) results,
as the IMSRG(3) truncation is the ``most complete''
IMSRG result we have available.

Considering the performance of the IMSRG(2) relative to the IMSRG(3),
we see that the difference in the correlation energy is about 1--2\% for most systems.
This also makes it clear how unusually good the IMSRG(2) results are
in the exceptional ${}^{16}\text{O}$ $NN$-only NAT case,
where the difference in the IMSRG(2) and IMSRG(3) results is closer to 0.1\%.
We also see that the IMSRG(2) results are systematically overbound
relative to the IMSRG(3) results.

Turning our attention to the IMSRG(3)-MP4 truncation,
we find that these results differ from the IMSRG(3) results by up to 1\%.
The results are also all less bound than the IMSRG(3) results,
making the IMSRG(2) and IMSRG(3)-MP4 results lower and upper bounds
on the IMSRG(3) result.
Considering that the IMSRG(3)-MP4 is the least computationally expensive approximate IMSRG(3) truncation we considered,
this provides a relatively cheap way to set a weak bound on where the IMSRG(3) result lands.
In the case where the many-body expansion converges systematically,
this bound should also encompass the effects of higher orders in the many-body expansion.

Turning our attention to the next two truncations,
the IMSRG(3)-$N^7$ and IMSRG(3)-$g^5$ truncations,
we find that the IMSRG(3)-$N^7$ results
are generally less bound than the IMSRG(3) results by about 0.5\%
(1\% in one case)
and the IMSRG(3)-$g^5$ results are generally more bound by about 0.1\%.
The gray bands in Fig.~\ref{fig:truncation_summary}
show the range of energies bounded by the results from these two truncations,
where we see that these bands always contain the IMSRG(3) results.
The IMSRG(3)-$N^7$ is of comparable expense and quality to the IMSRG(3)-MP4 truncation.
However, the IMSRG(3)-$g^5$ is considerably more expensive
and nearly as expensive as the full IMSRG(3).
This means that even once large-scale IMSRG(3)-MP4 and IMSRG(3)-$N^7$ are possible
IMSRG(3)-$g^5$ calculations may still be out of reach.
Still, if both IMSRG(3)-$N^7$ and IMSRG(3)-$g^5$ calculations are possible,
then these can be used to provide a robust bound on what the IMSRG(3) results could be.

\subsection{Performance for harder Hamiltonians}
\label{sec:hard_nn_results}

\begin{figure*}[p!]
    \centering
    \includegraphics{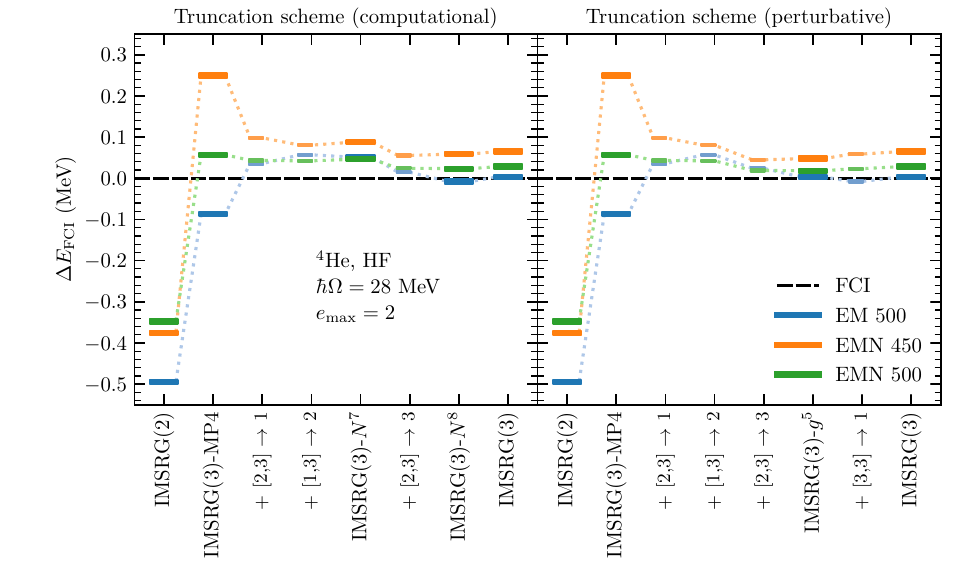}
    \caption{
        Differences of ground-state energies of ${}^{4}\text{He}$
        obtained in various truncation schemes to exact FCI results
        using several unevolved chiral Hamiltonians (see text for details) and an HF reference state
        following the computational (left panel)
        and perturbative (right panel) truncation ordering
        for the fundamental commutators.
        Thicker, darker bars correspond to the major truncations
        summarized in Table~\ref{tab:approximation_schemes}.
        Thinner, lighter bars
        correspond to intermediate truncations
        where a single fundamental commutator has been added
        relative to the truncation scheme to the left.
        \label{fig:he4_nn_all_hf_terms}
    }
\end{figure*}

\begin{figure*}[p!]
    \centering
    \includegraphics{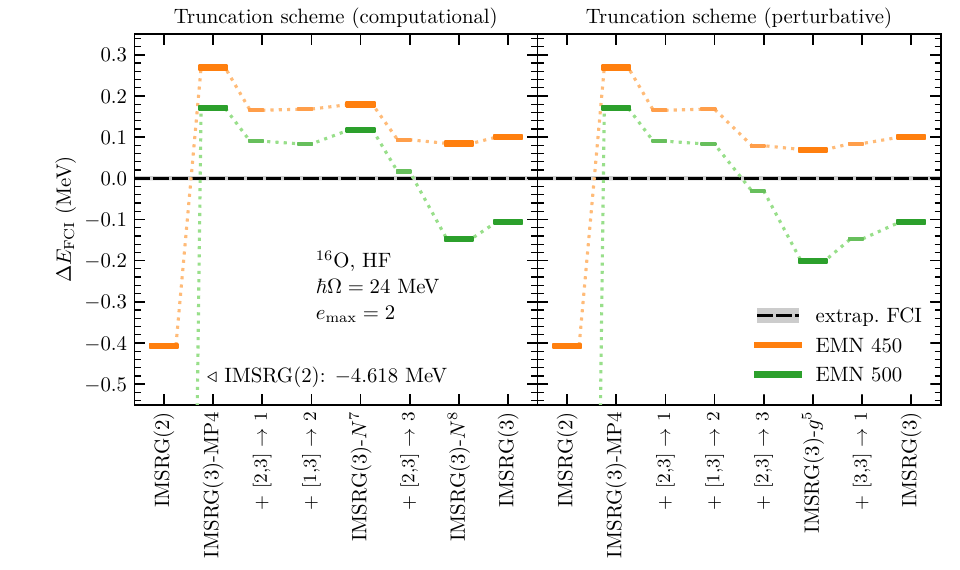}
    \caption{
        Same as Fig.~\ref{fig:he4_nn_all_hf_terms} but for ${}^{16}\text{O}$.
        \label{fig:o16_nn_all_hf_terms}
    }
\end{figure*}

In Fig.~\ref{fig:he4_nn_all_hf_terms},
we show the error to the exact FCI ground-state energy of ${}^{4}\text{He}$ for the harder $NN$-only Hamiltonians
for calculations using major and minor truncations schemes going from the IMSRG(2) approximation to the IMSRG(3) approximation.
The correlation energies for these Hamiltonians are about 8 to 10~MeV,
approximately double that of the EM 1.8 and EM 1.8/2.0 Hamiltonians in ${}^{4}\text{He}$.
We also note that the EM 500 Hamiltonian gives an unbound HF solution with a positive HF energy.

We see that for all three Hamiltonians
the IMSRG(2) overbinds the system substantially relative to the exact result.
These errors of about 350 to 500~keV correspond to errors of 3.5--5\% in the correlation energy.
The repulsive corrections from the IMSRG(3)-MP4 shift the obtained energies closer to the exact results.
Going from the IMSRG(3)-MP4 truncation
to the IMSRG(3)-$N^7$ and IMSRG(3)-$g^5$ truncations brings the IMSRG results within 100~keV of the exact results, a sub-1\% error in the correlation energy.
The higher-cost and higher-order corrections bring relatively small corrections,
and the final IMSRG(3) results remain within 100~keV of exact energies for all three Hamiltonians.
In Fig.~\ref{fig:o16_nn_all_hf_terms}, we show the results for ${}^{16}\text{O}$.
The approximate IMSRG(3) truncations systematically improve over the IMSRG(2),
and the final IMSRG(3) results differ from the exact results by just over 100~keV,
which is an error of about 0.5\% in the correlation energy for both Hamiltonians.
For the EM~500 Hamiltonian in the $e_{\text{max}}=2$ model space,
the IMSRG(2) calculation of ${}^{16}\text{O}$ does not converge.
The IMSRG(3) improves on this by delivering converged results
that differ from exact results by about 3\%,
stabilizing the solution of IMSRG flow equations.

We see that the IMSRG(3) offers substantial, systematic improvements over the IMSRG(2).
These improvements are largely already present
in approximate IMSRG(3) truncations with lower computational cost, such as the IMSRG(3)-$N^7$.
We note that the IMSRG(3) is not able to achieve as small of errors for these harder Hamiltonians as it is able to achieve for the EM 1.8 Hamiltonian with errors of up to 0.6\% in the correlation energy.
This suggests that the many-body expansion in the IMSRG converges more slowly when using harder Hamiltonians
(as one would also expect from perturbative arguments).
Still, the convergence behavior of the IMSRG many-body expansion is systematic in the cases discussed here, and the general trends discussed in Sec.~\ref{sec:truncation_performance} continue to hold.

\section{Summary and Outlook}
\label{sec:summary}

We performed the first systematic study of the inclusion of three-body operators
in the IMSRG in small model spaces.
To this end, we presented the fundamental commutators, the basic computational building blocks for the IMSRG,
required for the IMSRG(3) approximation
and introduced new truncations that include subsets of these commutators
to understand if one can reliably approximate the IMSRG(3).
We applied the full and approximate IMSRG(3) truncations
to the closed-shell ${}^{4}\text{He}$ and ${}^{16}\text{O}$
using $NN$-only and $NN$+$3N$ chiral Hamiltonians
with the Hartree-Fock and natural orbital single-particle bases.

When considering $NN$-only systems,
we compared the IMSRG(2) and IMSRG(3) results
to exact results in the same model space
obtained from FCI calculations for ${}^{4}\text{He}$
and from extrapolated FCI for ${}^{16}\text{O}$.
We found that the IMSRG(3) error to the (extrapolated) FCI correlation energy
was consistently about 0.3\% for the softest Hamiltonian considered
and up to 0.6\% for harder Hamiltonians.
Moreover,
the IMSRG(3) results improved systematically over the IMSRG(2) results,
where the error to the (extrapolated) FCI results varied quite significantly
for different bases and systems.
This suggests that the many-body expansion in the IMSRG,
which we have taken to the three-body-operator level in this work,
is well behaved.

We also considered the performance of various lower-cost approximate IMSRG(3) truncations
relative to the full IMSRG(3) approximation.
We used the perturbative analysis of Ref.~\cite{Herg16PR}
to investigate the expected size of contributions of terms
that are included in certain truncations and neglected in others.
We found that this perturbative analysis was able to
explain the size of contributions to the ground-state energy by individual terms
quite well.
As a result,
the energies calculated using approximate IMSRG(3) truncations
that included commutators based on their estimated perturbative importance
systematically converged to the full IMSRG(3) result.
The major truncation we considered in this approach, the IMSRG(3)-$g^5$,
reproduced the full IMSRG(3) results with very small errors
for both $NN$-only and $NN$+$3N$ Hamiltonians
across all frequencies, single-particle bases, and systems considered.

We also considered the organization of IMSRG(3) truncations
based on computational cost.
The key major truncation of this approach,
the IMSRG(3)-$N^7$,
has a lower computational cost than the IMSRG(3)-$g^5$ truncation.
The IMSRG(3)-$N^7$ truncation generally saw smaller errors relative to the full IMSRG(3)
than the IMSRG(2),
but the large contributions of missing commutators
prevented its performance from being as good as that of the IMSRG(3)-$g^5$ truncation.
The energy range given by the results from these two major IMSRG(3) truncation schemes (IMSRG(3)-$N^7$ and IMSRG(3)-$g^5$)
contained the full IMSRG(3) result in all of the cases we studied.

These IMSRG(3) approximations offer possibilities
for performing approximate IMSRG(3) calculations
where full IMSRG(3) calculations are no longer feasible and for studying the theoretical uncertainty due to the many-body truncation in IMSRG calculations.
The challenge going from here is the implementation
of full and approximate IMSRG(3) calculations for model spaces
where nuclear Hamiltonians are converged.
To achieve this, truncations in the three-body model space
will need to be imposed in addition to approximations to the IMSRG(3) truncation explored in this work.
In Ref.~\cite{Nova20PRCNeMgRch},
the natural orbitals are used to truncate the three-body model space in a way that accelerates convergence with respect to the employed model-space size.
The exploration of different three-body model-space truncations like this will be a key part of future work in the direction of reaching converged IMSRG(3) calculations.

\begin{acknowledgments}

We thank S.~R.~Stroberg for numerical checks to validate our implementation,
P.~Arthuis, H.~Hergert, S.~R.~Stroberg, and J.~M.~Yao for useful discussions,
and L.~Zurek for comments on the manuscript. This work was supported in part by the Deutsche Forschungsgemeinschaft (DFG, German Research Foundation) -- {Project-ID} 279384907 -- SFB 1245 and by the Max Planck Society.

\end{acknowledgments}

\onecolumngrid
\appendix

\section{IMSRG(3) flow equations}
\label{app:imsrg3_flow_equations}

The uncoupled (or $m$-scheme) IMSRG(3) flow equations are given by
\begin{align}
  \phantom{\frac{d\,W_{123456}}{ds}}
   & \begin{aligned}
    \mathllap{\frac{dE}{ds}} &=
    \sum_{pq}(n_p \bar{n}_q - \bar{n}_p n_q)\, \eta_{pq} f_{qp}
    + \frac{1}{4} \sum_{pqrs}(n_p n_q \bar{n}_r \bar{n}_s - \bar{n}_p \bar{n}_q n_r n_s)\,
    \eta_{pqrs} \Gamma_{rspq} \\
    &\quad + \frac{1}{36} \sum_{pqrstu}(n_p n_q n_r \bar{n}_s \bar{n}_t \bar{n}_u - \bar{n}_p \bar{n}_q \bar{n}_r n_s n_t n_u)\,
    \eta_{pqrstu} W_{stupqr}\,,
  \end{aligned}\label{eq:imsrg3_0body} \\
   & \begin{aligned}
    \mathllap{\frac{df_{12}}{ds}} &=
    \sum_{p}\left(\eta_{1p} f_{p2} - f_{1p} \eta_{p2}\right)
    + \sum_{pq}(n_p \bar{n}_q - \bar{n}_p n_q) \left(\eta_{pq} \Gamma_{1q2p} - f_{pq} \eta_{1q2p}\right)
    \\ & \quad
    + \frac{1}{2}\sum_{pqr}(\bar{n}_p \bar{n}_q n_{r} + n_p n_q \bar{n}_r) 
    \left(\eta_{1rpq} \Gamma_{pq2r} - \Gamma_{1rpq} \eta_{pq2r}\right)
    \\ & \quad
    + \frac{1}{4} \sum_{pqrs}(n_p n_q \bar{n}_r \bar{n}_s - \bar{n}_p \bar{n}_q n_r n_s)
    \left(\eta_{pqrs} W_{rs1pq2} - \Gamma_{pqrs} \eta_{rs1pq2}\right)
    \\ & \quad
    + \frac{1}{12} \sum_{pqrst}(n_p n_q n_r \bar{n}_s \bar{n}_t + \bar{n}_p \bar{n}_q \bar{n}_r n_s n_t)
    \left(\eta_{st1pqr} W_{pqrst2} - W_{st1pqr} \eta_{pqrst2} \right),
  \end{aligned}\label{eq:imsrg3_1body} \\
   & \begin{aligned}
    \mathllap{\frac{d\,\Gamma_{1234}}{ds}} &=
    (1 - P_{12}) \sum_{p} \left( \eta_{1p} \Gamma_{p234} - f_{1p} \eta_{p234}\right) 
    - (1 - P_{34}) \sum_{p} \left(\eta_{p3} \Gamma_{12p4} - f_{p3} \eta_{12p4}\right)
    \\ & \quad
    + \frac{1}{2}\sum_{pq}(\bar{n}_p \bar{n}_q - n_p n_q)
    \left(\eta_{12pq} \Gamma_{pq34} - \Gamma_{12pq} \eta_{pq34}\right)
    - (1 - P_{12})(1 - P_{34}) \sum_{pq}(n_p \bar{n}_q - \bar{n}_p n_q)\,
    \eta_{p23q} \Gamma_{1qp4}
    \\ & \quad
    + \sum_{pq}(n_p \bar{n}_q - \bar{n}_p n_q) \left(\eta_{pq} W_{12q34p} - f_{pq} \eta_{12q34p}\right)
    \\ & \quad
    + \frac{1}{2} (1 - P_{12}) \sum_{pqr}(\bar{n}_p \bar{n}_q n_r + n_p n_q \bar{n}_r)
    \left(\eta_{r1pq} W_{pq234r} - \Gamma_{r1pq} \eta_{pq234r}\right) 
    \\ & \quad
    - \frac{1}{2} (1 - P_{34}) \sum_{pqr}(\bar{n}_p \bar{n}_q n_r + n_p n_q \bar{n}_r) \left(\eta_{pqr3} W_{12rpq4} - \Gamma_{pqr3} \eta_{12rpq4}\right)
    \\ & \quad
    + \frac{1}{6} \sum_{pqrs}(\bar{n}_p \bar{n}_q \bar{n}_r n_s - n_p n_q n_r \bar{n}_s)
    \left(\eta_{12spqr} W_{pqr34s} - W_{12spqr} \eta_{pqr34s} \right)
    \\ & \quad
    + \frac{1}{4} (1 - P_{12}) (1 - P_{34}) \sum_{pqrs}(n_p n_q \bar{n}_r \bar{n}_s - \bar{n}_p \bar{n}_q n_r n_s)\,
     \eta_{pq1rs3} W_{rs2pq4}\,,
  \end{aligned}\label{eq:imsrg3_2body} \\
   & \begin{aligned}
    \mathllap{\frac{d\,W_{123456}}{ds}} &=
    P(12/3) P(45/6) \sum_{p} \left(\eta_{3p45} \Gamma_{126p} - \Gamma_{3p45} \eta_{126p}\right)
    \\ & \quad
    + P(12/3) \sum_{p} \left(\eta_{3p} W_{12p456} - f_{3p} \eta_{12p456}\right)
    - P(45/6) \sum_{p} \left(\eta_{p6} W_{12345p} - f_{p6} \eta_{12345p}\right)
    \\ & \quad
    + \frac{1}{2} P(12/3) \sum_{pq}(\bar{n}_p \bar{n}_q - n_p n_q)
    \left(\eta_{12pq} W_{pq3456} - \Gamma_{12pq} \eta_{pq3456}\right)
    \\ & \quad
    - \frac{1}{2}P(45/6) \sum_{pq}(\bar{n}_p \bar{n}_q - n_p n_q) 
    \left(\eta_{pq45} W_{123pq6} - \Gamma_{pq45} \eta_{123pq6}\right)
    \\ & \quad
    + P(12/3)P(45/6) \sum_{pq}(\bar{n}_p n_q - n_p \bar{n}_q)
    \left(\eta_{3pq6} W_{12q45p} - \Gamma_{3pq6} \eta_{12q45p}\right)
    \\ & \quad
    + \frac{1}{6} \sum_{pqr}(n_p n_q n_r + \bar{n}_p \bar{n}_q \bar{n}_r)
    \left(\eta_{123pqr} W_{pqr456} - W_{123pqr} \eta_{pqr456}\right)
    \\ & \quad
    + \frac{1}{2} P(12/3) P(45/6) \sum_{pqr}(\bar{n}_p \bar{n}_q n_r + n_p n_q \bar{n}_r)
    \left(\eta_{pq345r} W_{12rpq6} - W_{pq345r} \eta_{12rpq6}\right),
  \end{aligned}\label{eq:imsrg3_3body}
\end{align}
where the permutation operator $P_{pq}$ exchanges the indices $p$ and $q$ in the following expression. 
We further define the additional permutation operator
$P(pq/r) \equiv 1 - P_{pr} - P_{qr}$.
The action of the permutation operators in Eqs.~\eqref{eq:imsrg3_2body} and~\eqref{eq:imsrg3_3body}
ensures the antisymmetry of two- and three-body matrix elements over the course of the IMSRG evolution.
We note that the $m$-scheme IMSRG(3) flow equations agree with those in Ref.~\cite{Herg16PR}, except for the following typo:
\begin{enumerate}
    \item The occupation numbers in the term on the third row of Eq.~\eqref{eq:imsrg3_1body} are corrected.
\end{enumerate}
Our expressions differ somewhat because we do not use the Hermiticity of the Hamiltonian
and the anti-Hermiticity of the generator to manipulate the terms.
We note that there is no possible reduction in the computational cost
obtainable by these manipulations.
We also provide a list of corrections between our $m$-scheme
IMSRG(3) fundamental commutators and those in Ref.~\cite{Herg16PR}:
\begin{enumerate}
    \item Our expression for the \fcommtext{1}{3}{2} commutator has an overall factor of $1/4$ relative to that of Ref.~\cite{Herg16PR}.
    \item We include an additional term in the \fcommtext{2}{3}{3} commutator that was missing in Ref.~\cite{Herg16PR}.
    \item We provide an expression for the \fcommtext{2}{3}{2} commutator that is generally valid.
    The expression given in Ref.~\cite{Herg16PR} is valid only when one of $A$ and $B$ is Hermitian and the other is anti-Hermitian.
\end{enumerate}

\section{Spherical fundamental commutators}

In practice, the IMSRG(3) framework is applied to closed-shell systems with a spherical reference state.
Given the shared rotational symmetry of the reference state and nuclear Hamiltonians,
one can choose a spherical single-particle basis
and use angular-momentum-coupling techniques
to significantly reduce the storage and computational cost of the IMSRG(3) solution.

\subsection{Primer on angular-momentum coupling}

We offer a brief introduction to the concepts of angular-momentum coupling
and the associated notation.
For a more detailed treatment of the formalism of angular-momentum coupling,
we refer readers to Refs.~\cite{Vars88angmom,Suho07angmom}.

The single-particle basis is chosen to consist of spherical states
\begin{align}
    \ket{p} \equiv \ket{\xi_{p} j_p m_p} \equiv \ket{\tilde{p} m_p},
\end{align}
with the total angular momentum $j_p$,
the angular-momentum projection $m_p$,
and the remaining quantum numbers that characterize the state $\xi_p$.
In nuclear applications, $\xi = (n, l, t)$,
with the radial quantum number $n$,
the orbital angular momentum $l$,
and the isospin projection $t$.
The reduced single-particle index $\tilde{p}$
is a collective index for all the quantum numbers of the state besides $m_p$
and always has an associated $j_p$.
These spherical states are eigenstates of the one-body total angular momentum squared $J^2$
and the $z$ projection of the one-body total angular momentum $J_z$.

When using a spherical single-particle basis,
the one-body matrix elements of operators that are scalars under rotations in space and spin
(as is the case for the Hamiltonian and the generator in the IMSRG),
\begin{equation}
    \braket{\xi_p j_p m_p | O | \xi_q j_q m_q} = \braket{\tilde{p} m_p | O | \tilde{q} m_q},
\end{equation}
are diagonal in $j_p = j_q \equiv J_{O}$
and in $m_p = m_q \equiv M_{O}$
and independent of $M_{O}$.
This allows for the compact representation of the one-body matrix elements as
\begin{equation}
    O_{\tilde{p}\tilde{q}}^{J_{O}} \equiv \braket{\xi_p, j_p=J_{O}, m_p=j_p | O | \xi_q, j_q=J_{O}, m_q=j_q},
\end{equation}
where the single-particle indices now only run over reduced indices.
We have introduced a channel notation
where the superscript $J_{O}$ indicates that the matrix elements
are partitioned into channels
where matrix elements in each channel are nonzero only when $j_{p} = j_{q} = J_{O}$.
While it is conventional to use $j$, $J$, and $\mathcal{J}$
for one-, two-, and three-body angular momenta, respectively, 
we opt instead to use $j$ only for single-particle angular momenta
and $J$ for all angular momenta that appear in one-,
two-, and three-body angular-momentum channels.

The antisymmetric two-body states
\begin{equation}
    \ket{p q} \equiv \crea{p} \crea{q} \ket{0}
\end{equation}
may be coupled to two-body total angular momentum $J$
using the Clebsch-Gordan coefficients
\begin{equation}
    \clebsch{j_p}{m_p}{j_q}{m_q}{J}{M} = 
    \braket{(\tilde{p}\tilde{q})J M | p q},
\end{equation}
yielding the coupled two-body states
\begin{equation}
    \ket{(\tilde{p}\tilde{q})J M} = \sum_{m_p m_q}
    \clebsch{j_p}{m_p}{j_q}{m_q}{J}{M}
    \ket{p q},
\end{equation}
which are eigenstates of two-body $J^2$ and $J_z$.

When using coupled two-body states,
the two-body matrix elements of scalars under rotations in space and spin,
\begin{equation}
    \braket{(\tilde{p} \tilde{q}) J_{pq} M_{pq} | O | (\tilde{r} \tilde{s}) J_{rs} M_{rs}},
\end{equation}
are diagonal in $J_{pq} = J_{rs} \equiv J_{O}$
and in $M_{pq} = M_{rs} \equiv M_{O}$
and independent of $M_{O}$.
This allows for the compact representation of these coupled matrix elements as
\begin{equation}
    O_{\tilde{p}\tilde{q}\tilde{r}\tilde{s}}^{J_{O}} \equiv 
    \braket{(\tilde{p} \tilde{q}) J_{pq} = J_{O}, M_{pq} = J_{pq} | O | (\tilde{r} \tilde{s}) J_{rs} = J_{O}, M_{rs}= J_{rs}},
\end{equation}
where the single-particle indices again only run over reduced indices,
and the matrix elements have a channel structure 
that specifies to which total angular momentum $J_{O}$
the bra and ket states are coupled.

This approach is quickly extended to three-body states
\begin{equation}
    \ket{p q r} \equiv \crea{p} \crea{q} \crea{r} \ket{0},
\end{equation}
where the angular momenta $j_p$ and $j_q$ are coupled to an intermediate two-body angular momentum $J_{pq}$
that is then coupled with $j_r$ to the three-body angular momentum $J$,
yielding the coupled three-body states
\begin{equation}
    \ket{[(\tilde{p}\tilde{q})J_{pq}\tilde{r}] J M} =
    \sum_{m_p m_q M_{pq} m_r}
    \clebsch{j_p}{m_p}{j_q}{m_q}{J_{pq}}{M_{pq}}
    \clebsch{J_{pq}}{M_{pq}}{j_r}{m_r}{J}{M}
    \ket{p q r},
\end{equation}
which are eigenstates of the three-body $J^2$ and $J_z$.
Here, we made a choice to couple the $p$ and $q$ indices first and then the $r$ index.
One could also couple two different indices in the first coupling step and then couple the remaining index last
to arrive at valid eigenstates of $J^2$ and $J_z$.
One arrives at a similar representation for the coupled three-body matrix elements of a scalar operator,
\begin{equation}
    O_{\tilde{p}\tilde{q}\tilde{r}\tilde{s}\tilde{t}\tilde{u}}^{(J_{O}, J_{pq}, J_{st})}
    \equiv \braket{
        [(\tilde{p}\tilde{q})J_{pq}\tilde{r}] J_{pqr} = J_{O}, M_{pqr} = J_{pqr}
        | O |
        [(\tilde{s}\tilde{t})J_{st}\tilde{u}] J_{stu} = J_{O}, M_{stu} = J_{stu}
    },
\end{equation}
with $J_{O} = J_{pqr} = J_{stu}$ and $M_{pqr} = M_{stu}$.
The channel structure of three-body coupled matrix elements
is complicated by the appearance of the intermediate couplings
$J_{pq}$ and $J_{st}$,
which do not have to be equal.

Angular-momentum coupling allows one to reduce
the working equations of a theory
to expressions that depend only on the coupled matrix elements
discussed above.
The substantial reduction in storage requirements
due to working with coupled matrix elements
and in computational cost by having any purely geometric dependence
on angular-momentum projection analytically simplified
is essential to making IMSRG(3) calculations tractable.

For this work,
we used the automated angular-momentum-coupling tool \textsc{amc}~\cite{Tich20amc}
to generate coupled expressions for the fundamental commutators.
The generated expressions and their implementations were validated
by evaluating the coupled and uncoupled implementations
for the same input
and observing that the same coupled matrix elements were produced.

\subsection{Coupled expressions for fundamental commutators}

In the following,
we present the coupled expressions for the fundamental commutators
required for the IMSRG(3).
We drop the tilde from reduced single-particle indices,
as all matrix elements are coupled matrix elements,
and thus all indices on the matrix elements are reduced single-particle indices.

The expressions are nonantisymmetrized,
so the resulting two- and three-body coupled matrix elements
must be antisymmetrized by applying the appropriate antisymmetrizer to the bra and ket indices.
The antisymmetrization of two-body bra indices
is given by
\begin{equation}
    \bar{O}_{pqrs}^{J_{O}} \equiv \mathcal{A}_{2} O_{pqrs}^{J_{O}}
    = \frac{1}{2}
    \left[
        O_{pqrs}^{J_{O}}
        - (-1)^{j_p + j_q - J_{O}} O_{qprs}^{J_{O}}
    \right],
\end{equation}
where $\mathcal{A}_{2}$ is the two-body antisymmetrizer
and the output matrix elements $\bar{O}_{pqrs}^{J_{O}}$
are antisymmetric under exchange of $p$ and $q$.
If the input matrix elements $O_{pqrs}^{J_{O}}$ are already antisymmetric in $p$ and $q$,
the antisymmetrization does nothing and the input and output matrix elements are identical.
Similarly, the antisymmetrization of two-body ket indices is given by
\begin{equation}
    \bar{O}_{pqrs}^{J_{O}} \equiv  O_{pqrs}^{J_{O}} \mathcal{A}_{2}
    = \frac{1}{2}
    \left[
        O_{pqrs}^{J_{O}}
        - (-1)^{j_r + j_s - J_{O}} O_{pqsr}^{J_{O}}
    \right].
\end{equation}
The antisymmetrization of three-body bra indices is given by
\begin{align}
    \phantom{\bar{O}_{pqrstu}^{(J_{O}, J_{pq}, J_{st})} \equiv \mathcal{A}_{3} O_{pqrstu}^{(J_{O}, J_{pq}, J_{st})}}
    & \begin{aligned}
    \mathllap{\bar{O}_{pqrstu}^{(J_{O}, J_{pq}, J_{st})} \equiv \mathcal{A}_{3} O_{pqrstu}^{(J_{O}, J_{pq}, J_{st})}} &=\frac{1}{6}\Bigg[
    O_{pqrstu}^{(J_{O}, J_{pq}, J_{st})}
    + \hat{J}_{pq} \sum_{J_{2}} \hat{J}_{2}
    \sixj{j_p}{j_q}{J_{pq}}{j_r}{J_{O}}{J_{2}}
    O_{rqpstu}^{(J_{O}, J_{2}, J_{st})}
    \\ & \qquad
    - (-1)^{j_q + j_r - J_{pq}} \hat{J}_{pq}
    \sum_{J_{2}} (-1)^{J_{2}} \hat{J}_{2}
    \sixj{j_q}{j_p}{J_{pq}}{j_r}{J_{O}}{J_{2}}
    O_{prqstu}^{(J_{O}, J_{2}, J_{st})}
    \\ & \qquad
    - (-1)^{j_p + j_q - J_{pq}} \hat{J}_{pq}
    \sum_{J_{2}} \hat{J}_{2}
    \sixj{j_q}{j_p}{J_{pq}}{j_r}{J_{O}}{J_{2}}
    O_{rpqstu}^{(J_{O}, J_{2}, J_{st})}
    \\ & \qquad
    - (-1)^{j_q + j_r} \hat{J}_{pq}
    \sum_{J_{2}} (-1)^{J_{2}} \hat{J}_{2}
    \sixj{j_p}{j_q}{J_{pq}}{j_r}{J_{O}}{J_{2}}
    O_{qrpstu}^{(J_{O}, J_{2}, J_{st})}
    - (-1)^{j_p + j_q - J_{pq}}
    O_{qprstu}^{(J_{O}, J_{pq}, J_{st})}
    \Bigg]\,,
    \end{aligned}
\end{align}
with the three-body antisymmetrizer $\mathcal{A}_3$, $\hat{J} \equiv \sqrt{2 J + 1}$,
and the Wigner 6$j$ symbols
\begin{equation*}
    \sixj{j_1}{j_2}{j_3}{j_4}{j_5}{j_6}.
\end{equation*}
The antisymmetrization of three-body ket indices is given by
\begin{align}
    \phantom{\bar{O}_{pqrstu}^{(J_{O}, J_{pq}, J_{st})} \equiv  O_{pqrstu}^{(J_{O}, J_{pq}, J_{st})} \mathcal{A}_{3}}
    & \begin{aligned}
    \mathllap{\bar{O}_{pqrstu}^{(J_{O}, J_{pq}, J_{st})} \equiv O_{pqrstu}^{(J_{O}, J_{pq}, J_{st})} \mathcal{A}_{3}} &=\frac{1}{6}\Bigg[
    O_{pqrstu}^{(J_{O}, J_{pq}, J_{st})}
    + \hat{J}_{st} \sum_{J_{2}} \hat{J}_{2}
    \sixj{j_s}{j_t}{J_{st}}{j_u}{J_{O}}{J_{2}}
    O_{pqruts}^{(J_{O}, J_{pq}, J_{2})}
    \\ & \qquad
    - (-1)^{j_t + j_u - J_{st}} \hat{J}_{st}
    \sum_{J_{2}} (-1)^{J_{2}} \hat{J}_{2}
    \sixj{j_t}{j_s}{J_{st}}{j_u}{J_{O}}{J_{2}}
    O_{pqrsut}^{(J_{O}, J_{pq}, J_{2})}
    \\ & \qquad
    - (-1)^{j_s + j_t - J_{st}} \hat{J}_{st}
    \sum_{J_{2}} \hat{J}_{2}
    \sixj{j_t}{j_s}{J_{st}}{j_u}{J_{O}}{J_{2}}
    O_{pqrust}^{(J_{O}, J_{pq}, J_{2})}
    \\ & \qquad
    - (-1)^{j_t + j_u} \hat{J}_{st}
    \sum_{J_{2}} (-1)^{J_{2}} \hat{J}_{2}
    \sixj{j_s}{j_t}{J_{st}}{j_u}{J_{O}}{J_{2}}
    O_{pqrtus}^{(J_{O}, J_{pq}, J_{2})}
    - (-1)^{j_s + j_t - J_{st}}
    O_{pqrtsu}^{(J_{O}, J_{pq}, J_{st})}
    \Bigg]\,.
    \end{aligned}
\end{align}

\subsubsection{\texorpdfstring{$\fcomm{1}{1}{\circ}$}{11X commutators}}

\begin{align}
    \phantom{C_{12}^{J_C}}
    & \begin{aligned}
    \mathllap{C_{12}^{J_C}} &=
    \sum_{p}\left(
        A_{1p}^{J_C} B_{p2}^{J_C}
        - B_{1p}^{J_C} A_{p2}^{J_C}
    \right),
    \end{aligned}\label{eq:comm_111_coupled}\\
    & \begin{aligned}
    \mathllap{\op{C}{0}} &=
    \sum_{J_p} \hat{J}_{p}^2
    \sum_{pq} (n_p \bar{n}_q - \bar{n}_p n_q)
    A_{pq}^{J_p} B_{qp}^{J_p}\,.
    \end{aligned}\label{eq:comm_110_coupled}
\end{align}

\subsubsection{\texorpdfstring{$\fcomm{1}{2}{\circ}$}{12X commutators}}

\begin{align}
    \phantom{C_{1234}^{J_C}}
    & \begin{aligned}
    \mathllap{C_{1234}^{J_C}} &=
    2 \sum_{J_{A}} \sum_{p}
    \left(
        A_{1p}^{J_{A}} B_{p234}^{J_C}
        - A_{p3}^{J_A} B_{12p4}^{J_C}
    \right),
    \end{aligned}\label{eq:comm_122_coupled}\\
    & \begin{aligned}
    \mathllap{C_{12}^{J_C}} &=
    \frac{1}{\hat{J}_{C}^2} \sum_{J_B} \hat{J}_{B}^2 \sum_{J_A}
    \sum_{pq} (n_p \bar{n}_q - \bar{n}_p n_q)
    A_{pq}^{J_A} B_{1q2p}^{J_B}\,.
    \end{aligned}\label{eq:comm_121_coupled}
\end{align}

\subsubsection{\texorpdfstring{$\fcomm{2}{2}{\circ}$}{22X commutators}}

\begin{align}
    \phantom{C_{123456}^{(J_C, J_{12}, J_{45})}}
    & \begin{aligned}
    \mathllap{C_{123456}^{(J_C, J_{12}, J_{45})}} &=
    -9 \hat{J}_{12} \hat{J}_{45}
    \sum_{p}
    \sixj{j_3}{j_p}{J_{45}}{j_6}{J_C}{J_{12}}
    \left(
        A_{3p45}^{J_{45}}
        B_{126p}^{J_{12}}
        - B_{3p45}^{J_{45}}
        A_{126p}^{J_{12}} 
    \right),
    \end{aligned}\label{eq:comm_223_coupled}\\
    & \begin{aligned}
    \mathllap{C_{1234}^{J_C}} &=
    D_{1234}^{J_C} + E_{1234}^{J_C}\,,
    \end{aligned}\label{eq:comm_222_coupled_full} \\
    & \begin{aligned}
    \mathllap{D_{1234}^{J_C}} &=
    \frac{1}{2}
    \sum_{pq}(\bar{n}_p \bar{n}_q - n_p n_q)
    \left(
        A_{12pq}^{J_C} B_{pq34}^{J_C}
        - B_{12pq}^{J_C} A_{pq34}^{J_C}
    \right),
    \end{aligned}\label{eq:comm_222_coupled_term1} \\
    & \begin{aligned}
    \mathllap{\overline{E}_{1432}^{J_{C}^{\prime}}} &=
    4 \sum_{pq} (n_p \bar{n}_{q} - \bar{n}_p n_{q})
    \overline{A}_{p q 32}^{J_{C}^{\prime}}
    \overline{B}_{14pq}^{J_{C}^{\prime}}\,,
    \end{aligned}\label{eq:comm_222_coupled_term2} \\
    & \begin{aligned}
    \mathllap{C_{12}^{J_C}} &=
    \frac{1}{2} \frac{1}{\hat{J}_{C}^2}
    \sum_{J_{pq}} \hat{J}_{pq}^2
    \sum_{pqr} (\bar{n}_p \bar{n}_q n_r + n_p n_q \bar{n}_r)
    \left(
        A_{1rpq}^{J_{pq}} B_{pq2r}^{J_{pq}}
        - B_{1rpq}^{J_{pq}} A_{pq2r}^{J_{pq}}
    \right),
    \end{aligned}\label{eq:comm_221_coupled} \\
    & \begin{aligned}
    \mathllap{\op{C}{0}} &=
    \frac{1}{4}
    \sum_{J_{pq}} \hat{J}_{pq}^2
    \sum_{pq}
    (n_p n_q \bar{n}_r \bar{n}_s - \bar{n}_p \bar{n}_q n_r n_s)
    A_{pqrs}^{J_{pq}} B_{rspq}^{J_{pq}}\,,
    \end{aligned}\label{eq:comm_220_coupled}
\end{align}
where we split the \fcommtext{2}{2}{2} commutator
in Eq.~\eqref{eq:comm_222_coupled_full}
into two terms,
Eq.~\eqref{eq:comm_222_coupled_term1}
and Eq.~\eqref{eq:comm_222_coupled_term2}.
The matrix elements of $\op{A}{2}$ and $\op{B}{2}$ in Eq.~\eqref{eq:comm_222_coupled_term2}
(the $\overline{A}$ and $\overline{B}$ objects)
are obtained by a Pandya transformation~\cite{Pand56pandya},
\begin{equation}\label{eq:pandya_twobody}
    \overline{O}_{1432}^{J_{O}^{\prime}} \equiv
    - \sum_{J_O} \hat{J}_{O}^2
    \sixj{j_1}{j_4}{J_{O}^{\prime}}{j_3}{j_2}{J_{O}}
    O_{1234}^{J_{O}}\,.
\end{equation}
The Pandya transformation is its own inverse,
so the output Pandya-transformed matrix elements in Eq.~\eqref{eq:comm_222_coupled_term2}
($\overline{E}_{1432}^{J_{C}^{\prime}}$)
must be Pandya transformed again 
to arrive at the standard coupled matrix elements
($E_{1234}^{J_C}$)
that contribute in Eq.~\eqref{eq:comm_222_coupled_full}
to obtain the full \fcommtext{2}{2}{2} commutator result.

\subsubsection{\texorpdfstring{$\fcomm{1}{3}{\circ}$}{13X commutators}}

\begin{align}
    \phantom{C_{123456}^{(J_C, J_{12}, J_{45})}}
    & \begin{aligned}
    \mathllap{C_{123456}^{(J_C, J_{12}, J_{45})}} &=
    3 \sum_{J_A} \sum_{p}
    \left[
        A_{3p}^{J_A} B_{12p456}^{(J_C, J_{12}, J_{45})}
        - A_{p6}^{J_A} B_{12345p}^{(J_C, J_{12}, J_{45})}
    \right],
    \end{aligned}\label{eq:comm_133_coupled}\\
    & \begin{aligned}
    \mathllap{C_{1234}^{J_C}} &=
    \frac{1}{\hat{J}_{C}^2}
    \sum_{J_B} \hat{J}_{B}^2
    \sum_{J_A}
    \sum_{pq} (n_p \bar{n}_q - \bar{n}_p n_q)
    A_{pq}^{J_A}
    B_{12q34p}^{(J_B, J_{C}, J_{C})}\,.
    \end{aligned}\label{eq:comm_132_coupled}
\end{align}

\subsubsection{\texorpdfstring{$\fcomm{2}{3}{\circ}$}{23X commutators}}

\begin{align}
    \phantom{C_{123456}^{(J_C, J_{12}, J_{45})}}
    & \begin{aligned}
    \mathllap{C_{123456}^{(J_C, J_{12}, J_{45})}} &=
    D_{123456}^{(J_C, J_{12}, J_{45})}
    + E_{123456}^{(J_C, J_{12}, J_{45})}\,,
    \end{aligned}\label{eq:comm_233_coupled_full}\\
    & \begin{aligned}
    \mathllap{D_{123456}^{(J_C, J_{12}, J_{45})}} &=
    \frac{3}{2} \sum_{pq} (\bar{n}_p \bar{n}_q - n_p n_q)
    \left[
        A_{12pq}^{J_{12}}
        B_{pq3456}^{(J_{C}, J_{12}, J_{45})}
        - A_{pq45}^{J_{45}}
        B_{123pq6}^{(J_{C}, J_{12}, J_{45})}
    \right],
    \end{aligned}\label{eq:comm_233_coupled_term1}\\
    & \begin{aligned}
    \mathllap{E_{123456}^{(J_C, J_{12}, J_{45})}} &=
    9 \sum_{J_{A}, J_{B}, J_{qp}}
    (-1)^{J_{B} + J_{C}} \hat{J}_{A}^2 \hat{J}_{B}^2 \hat{J}_{qp}^2
    \sum_{pq} (\bar{n}_p n_q - n_p \bar{n}_q)(-1)^{j_3 + j_q}
    \\ & \quad \times
    \sixj{j_6}{j_3}{J_{qp}}{j_p}{j_q}{J_{A}}
    \sixj{J_{12}}{J_{45}}{J_{qp}}{j_p}{j_q}{J_{B}}
    \sixj{J_{qp}}{J_{12}}{J_{45}}{J_{C}}{j_6}{j_3}
    A_{3pq6}^{J_{A}}
    B_{12q45p}^{(J_{B}, J_{12}, J_{45})}\,,
    \end{aligned}\label{eq:comm_233_coupled_term2}\\
    & \begin{aligned}
    \mathllap{C_{1234}^{J_C}} &=
    -\frac{(-1)^{J_{C}}}{\hat{J}_{C}} \sum_{J_{pq}, J_{B}} \hat{J}_{pq} \hat{J}_{B}^2
    \sum_{pqr} (\bar{n}_p \bar{n}_q n_r + n_p n_q \bar{n}_r)
    \\ & \quad \times
    \left[
    (-1)^{j_1 + j_2}
    \sixj{j_2}{j_1}{J_{C}}{j_r}{J_{B}}{J_{pq}}
    A_{r1pq}^{J_{pq}} B_{pq234r}^{(J_{B}, J_{pq}, J_{C})}
    - (-1)^{j_3 + j_4}
    \sixj{j_4}{j_3}{J_{C}}{j_r}{J_{B}}{J_{pq}}
    A_{pqr3}^{J_{pq}} B_{12rpq4}^{(J_{B}, J_{C}, J_{pq})}
    \right],
    \end{aligned}\label{eq:comm_232_coupled}\\
    & \begin{aligned}
    \mathllap{C_{12}^{J_C}} &=
    \frac{1}{4} \frac{1}{\hat{J}_{C}^2} \sum_{J_{pq}, J_{B}} \hat{J}_{B}^2
    \sum_{pqrs} (n_p n_q \bar{n}_r \bar{n}_s -  \bar{n}_p \bar{n}_q n_r n_s)
    A_{pqrs}^{J_{pq}} B_{rs1pq2}^{(J_{B}, J_{pq}, J_{pq})}\,,
    \end{aligned}\label{eq:comm_231_coupled}
\end{align}
where we split the \fcommtext{2}{3}{3} commutator in Eq.~\eqref{eq:comm_233_coupled_full} into two terms.

\subsubsection{\texorpdfstring{$\fcomm{3}{3}{\circ}$}{33X commutators}}

\begin{align}
    \phantom{C_{123456}^{(J_C, J_{12}, J_{45})}}
    & \begin{aligned}
    \mathllap{C_{123456}^{(J_C, J_{12}, J_{45})}} &=
    D_{123456}^{(J_C, J_{12}, J_{45})}
    + E_{123456}^{(J_C, J_{12}, J_{45})}\,,
    \end{aligned}\label{eq:comm_333_coupled_full}\\
    & \begin{aligned}
    \mathllap{D_{123456}^{(J_C, J_{12}, J_{45})}} &=
    \frac{1}{6} \sum_{J_{pq}} \sum_{pqr} (n_p n_q n_r + \bar{n}_p \bar{n}_q \bar{n}_r)
    \left[
    A_{123pqr}^{(J_C, J_{12}, J_{pq})}
    B_{pqr456}^{(J_C, J_{pq}, J_{45})}
    -
    B_{123pqr}^{(J_C, J_{12}, J_{pq})}
    A_{pqr456}^{(J_C, J_{pq}, J_{45})}
    \right],
    \end{aligned}\label{eq:comm_333_coupled_term1}\\
    & \begin{aligned}
    \mathllap{\overline{E}_{126453}^{(J_{C}^{\prime}, J_{12}, J_{45})}} &=
    \frac{9}{2} \sum_{J_{pq}} \sum_{pqr} (\bar{n}_p \bar{n}_q n_{r} + n_p n_q \bar{n}_{r})
    \Big[
    \overline{A}_{pqr453}^{(J_{C}^{\prime}, J_{pq}, J_{45})}
    \overline{B}_{126pqr}^{(J_{C}^{\prime}, J_{12}, J_{pq})}
    -
    \overline{B}_{pqr453}^{(J_{C}^{\prime}, J_{pq}, J_{45})}
    \overline{A}_{126pqr}^{(J_{C}^{\prime}, J_{12}, J_{pq})}
    \Big],
    \end{aligned}\label{eq:comm_333_coupled_term2}\\
    & \begin{aligned}
    \mathllap{C_{1234}^{J_C}} &=
    D_{1234}^{J_C} + E_{1234}^{J_C}\,,
    \end{aligned}\label{eq:comm_332_coupled_full}\\
     & \begin{aligned}
    \mathllap{D_{1234}^{J_C}} &=
    \frac{1}{6} \frac{1}{\hat{J}_{C}^2}
    \sum_{J_{pqr}} \hat{J}_{pqr}^2 \sum_{J_{pq}}
    \sum_{pqrs} (\bar{n}_p \bar{n}_q \bar{n}_r n_s - n_p n_q n_r \bar{n}_s)
    \left[
    A_{12spqr}^{(J_{pqr}, J_{C}, J_{pq})}
    B_{pqr34s}^{(J_{pqr}, J_{pq}, J_{C})}
    - B_{12spqr}^{(J_{pqr}, J_{C}, J_{pq})}
    A_{pqr34s}^{(J_{pqr}, J_{pq}, J_{C})}
    \right],
    \end{aligned}\label{eq:comm_332_coupled_term1}\\
     & \begin{aligned}
    \mathllap{E_{1234}^{J_C}} &=
    - (-1)^{j_1 + j_3 + J_{C}}
    \sum_{J_{A},J_{B}} \hat{J}_{A}^2 \hat{J}_{B}^2 (-1)^{J_{A} + J_{B}}
    \sum_{J_{pq}, J_{rs}} \sum_{J_{2}} \hat{J}_{2}^2
    \sum_{pqrs} (n_p n_q \bar{n}_r \bar{n}_s - \bar{n}_p \bar{n}_q n_r n_s)
    \\ & \quad \times
    \sixj{J_{pq}}{J_{rs}}{J_{2}}{j_3}{j_1}{J_{A}}
    \sixj{J_{pq}}{J_{rs}}{J_{2}}{j_2}{j_4}{J_{B}}
    \sixj{j_3}{j_4}{J_{C}}{j_2}{j_1}{J_{2}}
    A_{pq1rs3}^{(J_{A}, J_{pq}, J_{rs})}
    B_{rs2pq4}^{(J_{B}, J_{rs}, J_{pq})}
    \,,
    \end{aligned}\label{eq:comm_332_coupled_term2}\\
    & \begin{aligned}
    \mathllap{C_{12}^{J_C}} &=
    \frac{1}{12} \frac{1}{\hat{J}_{C}^2}
    \sum_{J_{pqr}, J_{pq}, J_{st}} \hat{J}_{pqr}^2
    \sum_{pqrst}
    (n_p n_q n_r \bar{n}_s \bar{n}_t
    + \bar{n}_p \bar{n}_q \bar{n}_r n_s n_t)
    \left[
        A_{st1pqr}^{(J_{pqr}, J_{st}, J_{pq})}
        B_{pqrst2}^{(J_{pqr}, J_{pq}, J_{st})}
        - B_{st1pqr}^{(J_{pqr}, J_{st}, J_{pq})}
        A_{pqrst2}^{(J_{pqr}, J_{pq}, J_{st})}
    \right],
    \end{aligned}\label{eq:comm_331_coupled} \\
    & \begin{aligned}
    \mathllap{\op{C}{0}} &=
    \frac{1}{36}
    \sum_{J_{pqr}, J_{pq}, J_{st}} \hat{J}_{pqr}^2
    \sum_{pqrstu}
    (n_p n_q n_r \bar{n}_s \bar{n}_t \bar{n}_u
    - \bar{n}_p \bar{n}_q \bar{n}_r n_s n_t n_u)
    A_{pqrstu}^{(J_{pqr}, J_{pq}, J_{st})}
    B_{stupqr}^{(J_{pqr}, J_{st}, J_{pq})}\,.
    \end{aligned}\label{eq:comm_330_coupled}
\end{align}
Here we split the \fcommtext{3}{3}{3} commutator in Eq.~\eqref{eq:comm_333_coupled_full} and the \fcommtext{3}{3}{2} commutator in Eq.~\eqref{eq:comm_332_coupled_full}
each into two terms.
The matrix elements of $\op{A}{3}$ and $\op{B}{3}$ in Eq.~\eqref{eq:comm_333_coupled_term2}
(the $\overline{A}$ and $\overline{B}$ objects)
are obtained by the three-body analog of the Pandya transformation,
\begin{equation}
    \overline{O}_{126453}^{(J_{O}^{\prime}, J_{12}, J_{45})} \equiv
    - \sum_{J_{O}} \hat{J}_{O}^2
    \sixj{J_{12}}{j_6}{J_{O}^{\prime}}{J_{45}}{j_3}{J_{O}}
    O_{123456}^{(J_{O}, J_{12}, J_{45})}\,.
\end{equation}
The output Pandya-transformed matrix elements in Eq.~\eqref{eq:comm_333_coupled_term2}
must be Pandya transformed again to arrive at the standard matrix elements
that contribute in Eq.~\eqref{eq:comm_333_coupled_full} to obtain the full \fcommtext{3}{3}{3} commutator result.

\twocolumngrid

\bibliography{strongint}

\end{document}